\documentclass[sigconf,nonacm,prologue,table]{acmart}
\usepackage{graphicx}
\usepackage{multirow} 
\usepackage{booktabs,threeparttable}
\usepackage{subcaption}
\usepackage{makecell}
\usepackage{listings}
\usepackage[switch]{lineno} 
\usepackage{xcolor}

\definecolor{lightgray}{rgb}{0.95, 0.95, 0.96}

\newcommand\vldbdoi{XX.XX/XXX.XX}

\newcommand\vldbavailabilityurl{URL_TO_YOUR_ARTIFACTS}

 \newcommand{\Kazem}[1]{\textcolor{blue}{\iffalse #1 \fi}}

\usepackage{tikz}
\usetikzlibrary{shapes.geometric}

\usepackage[linesnumbered,ruled]{algorithm2e}

\newcommand{\circlednumber}[1]{%
  \tikz[baseline]\node[circle,draw=black,fill=black,text=white,inner sep=1pt,minimum size=1pt,anchor=base] {#1};%
}
\newcommand{\DTT}{TDT}

\begin{document}

\title{Floating-Point Data Transformation for Lossless Compression }

\author{Samirasadat Jamalidinan}
\affiliation{%
  \institution{Electrical and Computer Engineering\\ McMaster University}
  \streetaddress{P.O. Box 1212}
  \city{Hamilton}
  \state{Canada}
  \postcode{43017-6221}
}
\email{jamalids@mcmaster.ca}

\author{Kazem Cheshmi}
\orcid{0000-0002-1825-0097}
\affiliation{%
  \institution{Electrical and Computer Engineering\\ McMaster University}
  \streetaddress{1 Th{\o}rv{\"a}ld Circle}
  \city{Hamilton}
  \country{Canada}
}
\email{cheshmi@mcmaster.ca}





\begin{abstract}
Floating-point data is widely used across various domains. Depending on the required precision, each floating-point value can occupy several bytes. Lossless storage of this information is crucial due to its critical accuracy, as seen in applications such as medical imaging and language model weights. In these cases, data size is often significant, making lossless compression essential. Previous approaches either treat this data as raw byte streams for compression or fail to leverage all patterns within the dataset. However, because multiple bytes represent a single value and due to inherent patterns in floating-point representations, some of these bytes are correlated. To leverage this property, we propose a novel data transformation method called Typed Data Transformation (\DTT{}) that groups related bytes together to improve compression. We implemented and tested our approach on various datasets across both CPU and GPU. \DTT{} achieves a geometric mean compression ratio improvement of 1.16$\times$ over state-of-the-art compression tools such as zstd, while also improving both compression and decompression throughput by 1.18--3.79$\times$.
\end{abstract}

\maketitle




\begin{figure}
    \centering
    \includegraphics[width=0.8\linewidth]{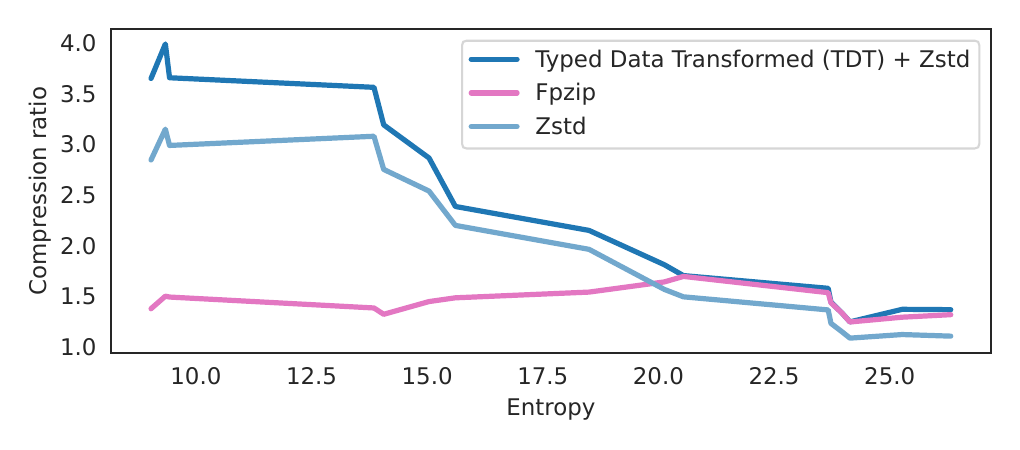}
    \caption{Typical compression behavior of floating-based compression tools such as fpzip and generic compression tools such as Zstd and how \DTT{} provides best of both worlds when combined with generic compression tools. }
    \label{fig:genericvsfloat}
\end{figure}
\section{Introduction}

Floating-point data is prevalent in numerous domains, including scientific computing~\cite{kissmann2016colliding,thoman2020rtx} and machine learning~\cite{touvron2023llama}. Compressing this data offers significant advantages, primarily by reducing storage space and memory footprint. In many applications involving floating-point values, lossless compression is crucial. This is particularly true in sensitive areas such as medical imaging, where even minor data alterations can have significant diagnostic implications, and in machine learning models, where preserving the precision of weights and biases is essential for model accuracy. This paper focuses on the lossless compression of floating-point datasets.

General-purpose lossless compression methods such as snappy~\cite{GoogleSnappy}, zstd~\cite{FacebookZstd}, and lz4~\cite{YannColletLZ4} treat all data, including floating-point values, as raw byte sequences for compression. However, typed data, such as floating-point numbers, is composed of multiple bytes that are often correlated, as they collectively represent a single value adhering to a specific standard. For instance, the IEEE 754 standard~\cite{IEEE754} is commonly used for floating-point representation, defining various type widths and allocating bytes for the exponent and mantissa. These components are themselves represented as integer or floating-point values. These standards inherently enforce partial patterns across floating-point values. However, these patterns are not explicitly exploited by generic compression tools because the data is treated simply as a stream of bytes.

Floating-point-based compression tools often capitalize on the unique properties of floating-point types for enhanced compression. These methods typically predict the next value based on prior values~\cite{liakos2022chimp,pelkonen2015gorilla} or a learned model~\cite{claggett2018spdp,lindstrom2006fast}. They then store only the differing bits, often using an XOR operation, which leverages the smooth transitions commonly observed in floating-point data streams. For instance, Gorilla~\cite{pelkonen2015gorilla} and Chimp~\cite{liakos2022chimp} exploit the frequent similarity of leading and trailing bits in consecutive floating-point values, making them highly compressible. While effective at saving space, these approaches often overlook the global patterns across bytes that generic compression methods leverage.
Figure~\ref{fig:genericvsfloat} illustrates this by comparing fpzip~\cite{lindstrom2006fast}, a floating-point-specific compression tool, with zstd~\cite{FacebookZstd}, a generic compression tool, across real-world floating-point datasets. As the figure shows, zstd effectively exploits the low-entropy properties of these datasets, while fpzip performs better on high-entropy datasets. This comparison highlights a performance gap between these two classes of methods.


\begin{figure}
    \centering
    \begin{subfigure}[b]{0.9\linewidth}
        \includegraphics[width=\linewidth]{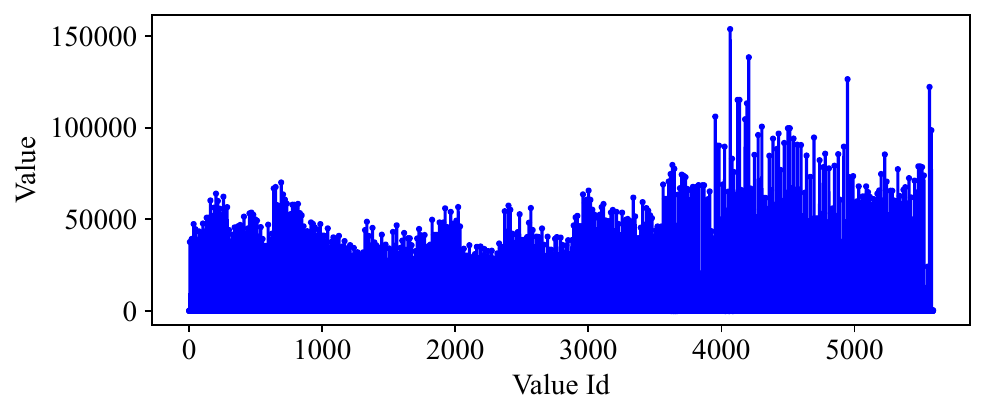}
        \caption{Dataset values}
        \label{fig:motiv_values}
    \end{subfigure}%
    
    \begin{subfigure}[b]{0.95\linewidth}
        \includegraphics[width=\linewidth]{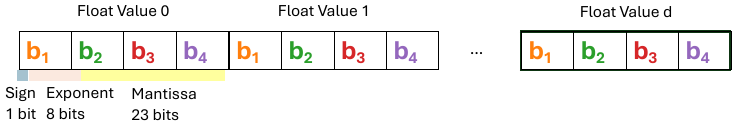}
        \caption{Original dataset}
        \label{fig:motiv_layoutieee}
    \end{subfigure}
    
    \begin{subfigure}[b]{0.9\linewidth}
        \includegraphics[width=\linewidth]{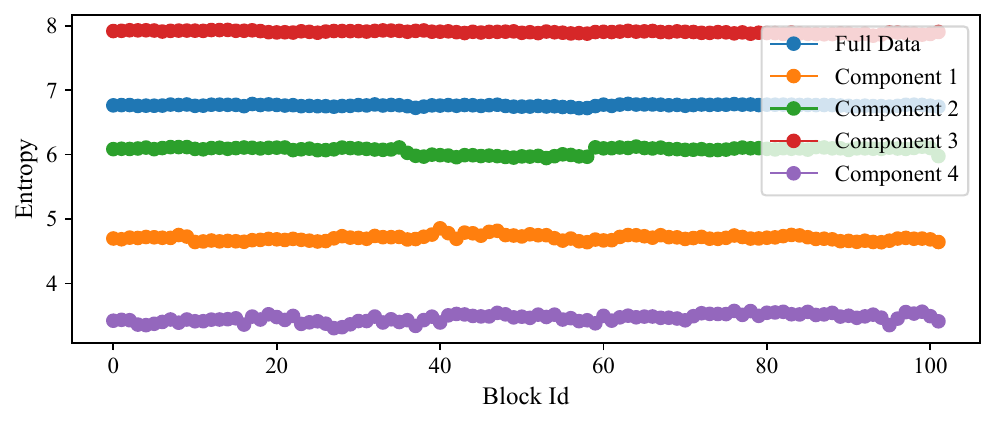}
        \caption{Entropy per block of dataset}
        \label{fig:motiv_entropy}
    \end{subfigure}
    
    \begin{subfigure}[b]{0.87\linewidth}
        \includegraphics[width=\linewidth]{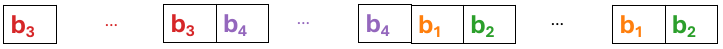}
        \caption{Transformed dataset}
        \label{fig:motiv_transformed}
    \end{subfigure}
    
    \begin{subfigure}[b]{0.9\linewidth}
        \includegraphics[width=\linewidth, height=4cm]{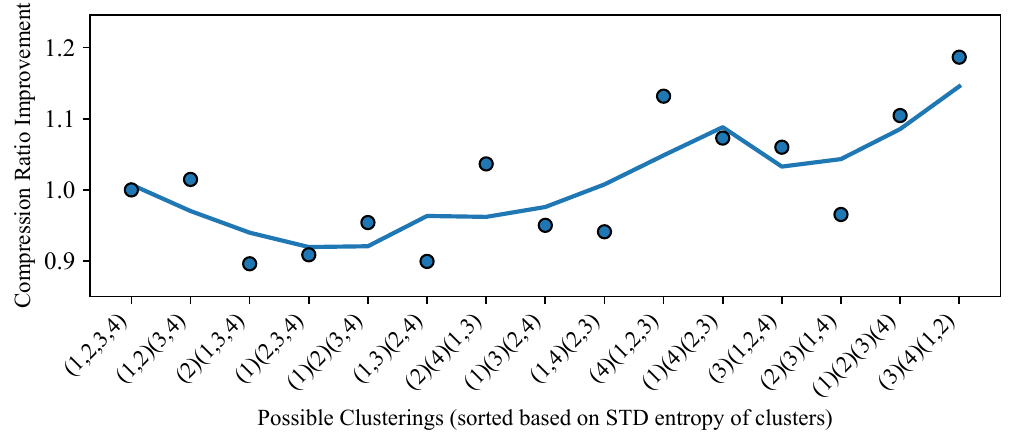}
        \caption{Data transformation space}
        \label{fig:motive2clustering}
    \end{subfigure}
    \caption{a) Range of values in a floating-point dataset. b) The byte representation of the dataset in memory. Four bytes store a float value, i.e., single precision, using the IEEE 754 standard.  c) The entropy of the dataset and each byte across the dataset, i.e., components. d) The byte representation of the dataset after applying \DTT{} transformation. e) All possible byte clustering for the dataset and their compression ratios. }
    \label{fig:motiv}
\end{figure}


Data transformation is a known technique for improving compression ratios. A data transform rearranges a dataset to expose more frequent patterns, thereby facilitating the application of other compression methods. While a data transformation does not alter the fundamental entropy of the data, it makes patterns easier to identify, which significantly aids compression methods such as dictionary coding. The Burrows-Wheeler Transform (BWT)~\cite{schindler1997fast}, for example, is commonly applied to string datasets, such as genomic data, to generate more repetitive patterns.
However, data transformation for floating-point datasets remains less explored. Shuffle and Bitshuffle~\cite{masui2015compression} are notable examples of data transforms specifically applied to floating-point datasets. These methods reorder  data before a subsequent compression technique is applied. Their key limitation, however, is their uniform treatment of all data, reordering bits and bytes identically across different applications, even though the relationship between bits and bytes can vary depending on data types and specific application contexts.


This paper proposes Typed Data Transformation (\DTT{}), a novel data transformation technique designed to enhance the compression of floating-point data by enabling generic compression tools to leverage floating-point characteristics. \DTT{} employs a new clustering technique to group correlated bytes within a dataset and then reorders these bytes to ensure that related information is stored contiguously. Also, \DTT{} facilitates independent compression and decompression of bytes, which improves load balance and, hence, throughput.
We conducted an extensive evaluation of this method across datasets from High-Performance Computing (HPC), Observation (OBS), database transactions (DB), and time-series (TS) domains. The methodology was also tested with a wide range of generic compression techniques, including lz4, zstd, zlib, snappy, bzip, and nvCOMP. Across all tested datasets, \DTT{} achieved a geometric mean (GMean) compression ratio of 1.86$\times$, outperforming the GMean of 1.62$\times$ achieved by the baseline compression methods. Additionally, \DTT{} inherently creates more parallelizable workloads, thereby amortizing the cost of transformation on parallel resources and significantly improving overall compression and decompression throughput.  

\begin{figure}
    \centering
    \includegraphics[width=1\linewidth]{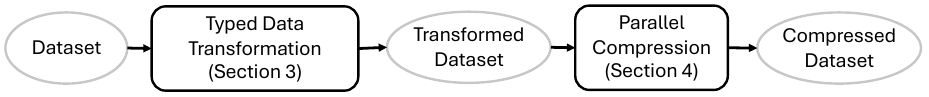}
    \caption{Overall view of decomposed typed data transformation (\DTT) and its application to de/compression}
    \label{fig:overall}
\end{figure}
\section{Motivation and Overview}
This section motivates how \DTT{} groups bytes using an example and provides an overview of the designed framework. 
A part of the \textit{solar\_wind} dataset from TS is shown in Figure~\ref{fig:motiv_values}. Each sample is shown with a single-precision floating-point type represented using the IEEE 754 standard. Figure~\ref{fig:motiv_layoutieee} also shows how the float values are laid out in memory using the IEEE standard. The single-precision (\texttt{float32}) format uses 32 bits divided into three fields: a 1-bit sign ($s$), an 8-bit exponent ($e$), and a 23-bit mantissa ($m$). In this format, each floating-point value $v$ is represented as $v = (-1)^s \times 2^{e - bias} \times (1 + m/2^{23})$, where $bias$ is 127.

%


The entropy of each byte often differs in floating-point values due to their underlying storage standard and common data patterns found in scientific and machine learning datasets. A higher entropy value indicates a more random data distribution and, thus, lower compressibility.
Figure~\ref{fig:motiv_entropy} illustrates the entropy of each byte (component) within the \textit{solar\_wind} dataset, alongside the dataset's overall entropy. As the figure shows, the entropy of individual bytes varies significantly from the dataset's total entropy. Specifically, components 1, 2, and 4 exhibit lower entropy than the dataset's overall entropy, while the third byte has higher entropy. The dataset's overall entropy is calculated assuming each symbol is one byte long. This non-uniform distribution of entropy can be used to reorganize the data, reducing the search space for compression, rather than simply looking for similar entropy components across the dataset.


Figure~\ref{fig:motiv_transformed} illustrates how the components are rearranged based on clustering entropy. Bytes 1 and 2 have more similarity; placing them contiguously creates data with more uniform entropy. Bytes 3 and 4 are distinct and should be separated. After the reordered data is created, bytes (components) with close entropies are formed to improve compressibility by a compression method. For example, using zstd, as an efficient compression method, to compress the original dataset and the transformed dataset leads to a compression ratio of 1.44$\times$ and 1.71$\times$, respectively.

Deciding how to cluster bytes is challenging as the number of clusters is not known, and finding interaction between bytes are data dependent.  
Figure~\ref{fig:motive2clustering} shows all possible clustering for four bytes and their associated compression ratios. The x-axis is sorted based standard deviation (STD) of entropies in components to show how non-uniformity correlates with compression ratios.  
To address the discussed challenges in data transformation, this paper proposed \DTT{} framework as shown in Figure~\ref{fig:overall}. 
\DTT{} formulates the data transformation into a byte clustering problem based on the predictability of some bytes in floating point data. 
As shown, \DTT{} reorders the dataset based on a byte clustering and packs it as a transformed dataset. The transformed dataset is blocked and compressed in parallel using existing compression methods.  
We discuss the details of \DTT{} and its clustering approach in Section~\ref{sec:dtransform} and show how it fits into a compression pipeline in Section~\ref{sec:compression}.

\begin{algorithm} [tb]
\SetAlgoLined
\DontPrintSemicolon
\SetAlgoLined
\begin{small}
\caption{\label{alg:clustering}\DTT{} Algorithm}
\SetKwInOut{Input}{Input}
\SetKwInOut{Output}{Output}
\Input{\,$D$ 
}
\Output{\,$P$}
\tcc{Step I: Feature Extraction (Section~\ref{sec:feature} )}
$X \leftarrow featureExtraction(D)$\; \label{lin:feature}
\tcc{Step II: Clustering (Section~\ref{sec:clustering})}
$L \leftarrow Linkage(X)$\; \label{lin:linkage}
$n \leftarrow size(X)$\;
\For{$b \leftarrow 1$ to $n$}{ \label{lin:clusterb}
    $C_b \leftarrow clustering(b, L)$\;
    $S_b \leftarrow score(C_b, X)$\;
}\label{lin:clustere}
$ m \leftarrow \text{argmax}(S)$\;\label{lin:scores}
$\mathcal{C} = C_m$\;\label{lin:final}
\tcc{Step III: Packing and Decomposition (Section~\ref{sec:packing})}
\For{ $c_i \in \mathcal{C}$ }{ \label{lin:packb}
    $P_i \leftarrow packing(D[c_i]) $\;
}\label{lin:packe}
\end{small}
\end{algorithm}

\section{Typed Data Transformation (\DTT)}
\label{sec:dtransform}

This section explains how \DTT{} is applied to the input data as shown in Figure~\ref{fig:overall}. We first provide an overview of the \DTT{} algorithm and then discuss the three steps of the algorithm. 
%

\subsection{Overview}

The goal of \DTT{} is to reorder data so that ``similar data'' segments are adjacent, thereby minimizing higher-order entropy. We define similar data segments when segments have comparable entropy or frequency characteristics. 
While this reordering doesn't change the overall entropy, it significantly benefits dictionary coders by reducing higher-order entropy, such as H1 entropy~\cite{manzini2001analysis}. Higher-order entropy finds deeper connections by using the frequency of multiple symbols, showing the potential of new patterns. Additionally, when treating these data segments with distinct entropies separately, it affects the entropy and thus helping entropy coders as well. Both of these compressors are common in generic compression tools. However, there are $d!$ possibilities of reordering a dataset with $d$ elements. This makes the problem to be NP hard problem.

\DTT{} uses a clustering approach to group similar data based on the observation that bytes in the same location of a float value follow a similar entropy value. 
By limiting the definition to groups of bytes in the same byte location in floating values, the search space reduces significantly to $n!$ possibility of clustering, where $n$ is the number of bytes in the float type. Considering possible reordering within clusters, clustering with reduced search space is still challenging.



Algorithm~\ref{alg:clustering} shows that \DTT{} takes dataset $D$ as input and produces reordered dataset $P$ in three steps to put bytes with similar entropies next to each other. 
%
The input dataset $D$ is shown as a list of float values $(b_1, b_2, \dots, b_n)$ where $b_i$ is the $i^{th}$ byte in value in $D$ and $n$ is the word width of the type in the dataset.  
The first step of the \DTT{} algorithm extracts features from each byte locations of $D$ and creates feature vector $X = (x_1, x_2, \dots, x_n)$, where each $x_i$ is a feature vector of byte $b_i$ (e.g., a vector in a multi-dimensional space). Then the second step clusters the features to create the clustering $\mathcal{C} = (C_1, C_2, \dots, C_k)$ where $C_j$ is a byte cluster $j$ including a set of features $x_i$ corresponding to byte IDs $b_i$. Finally, the third step creates the reordered dataset by storing bytes of the same cluster next to each other and in a separate stream as a list of $P$. The details of these three steps are explained in the rest of this section.





\begin{table}[!ht]
\centering
\caption{Selected Features for the clustering algorithm. $H_j$: Entropy of the $j$-th block as shown in Equation~\ref{eq:entropy}. $N$: Total number of blocks. $\text{Count}(i)$: Number of occurrences of byte value $i$ in the entire byte group.}
\resizebox{\linewidth}{!}{%
\begin{tabular}{lp{2.6cm}l}
\toprule
& \textbf{Feature Name} & \textbf{Mathematical Formulation} \\
\midrule
\rowcolor{lightgray} \multirow{4}{*}{} & Average Entropy & $\frac{1}{N} \sum_{j=1}^{N} H_j$ \\
& Standard Deviation of Entropy & $\sqrt{\frac{1}{N-1} \sum_{j=1}^{N} (H_j - \text{Average Entropy})^2}$ \\
\rowcolor{lightgray} & Maximum Entropy & $\max(H_1, H_2, \dots, H_N)$ \\
& Minimum Entropy & $\min(H_1, H_2, \dots, H_N)$ \\
\hline
\rowcolor{lightgray} & Normalized Byte Frequency & $f_i = \frac{\text{Count}(i)}{\sum_{j=0}^{255} \text{Count}(j)}$, $i = 0, 1, \dots, 255$ \\
\bottomrule
\end{tabular}%
}
\label{tab:feature}
\end{table}

\subsection{Step I: Feature Extraction}
\label{sec:feature}
While the entropies of bytes are uniform but their value changes across datasets. Therefore, using one entropy for one component across datasets does not represent it well for data transformation. 
To capture the complex characteristics of data for compression, the first step of the \DTT{} algorithm uses a combination of statistical features based on entropy to calculate the similarity between bytes.
The feature extraction process is designed to capture the statistical characteristics of each byte group over a fixed-size block. Compressing a block of data is common due to limited fast memory in computers. The $featureExtraction$ function in line~\ref{lin:feature} in Algorithm~\ref{alg:clustering} computes a set of entropy-based features from the input dataset.

The five features selected to represent information within a block of bytes are shown in Table~\ref{tab:feature}. The first four features are based on entropy $H$, as defined in Equation~\ref{eq:entropy}.
\begin{equation}
    H = - \sum_{i=1}^{q} p_i \log_2(p_i)
    \label{eq:entropy}
\end{equation}
where $p_i$ is the probability of the $i$-th byte value within the window, and $q$ is the total number of unique byte values (256 for 8-bit data). These four features—average entropy, standard deviation of entropy, maximum entropy, and minimum entropy— capture, respectively, overall randomness, inconsistencies in data randomness, and the maximum and minimum variability within the data.
We also use normalized byte frequency as the only feature in the frequency category, as shown in Table~\ref{tab:feature}. The normalized byte frequency of each byte value (0--255) is calculated for the entire byte group. This 256-dimensional vector captures the distribution of byte occurrences and serves as a global signature for the group.

The selected feature by the algorithm is a combination of all five features listed in Table~\ref{tab:feature}, often depending on the design choice, such as compression method and dataset properties.
 The extracted features for each byte group include the entropy-based metrics (average, standard deviation, maximum, and minimum entropy) and the byte frequency vector. These features are concatenated into a single feature vector, providing a comprehensive statistical representation of the byte group as shown with \( x_i \in \mathbb{R}^{260} \) as for the feature vector of the \(i\)th byte group:  
\[
x_i = \left[ \bar{H}_i,\, \sigma(H_i),\, \max(H_i),\, \min(H_i),\, f_i(0),\, f_i(1),\, \dots,\, f_i(255) \right],
\]
where components are defined based on Table~\ref{tab:feature} for a given byte group $i$. 

\subsection{Step II. Clustering}
\label{sec:clustering}
The clustering problem in the second step of the \DTT{} algorithm takes a list of $n$ features $X = (x_1, x_2, \dots, x_n)$ and partitions them into $k$ clusters $\mathcal{C}$. The goal is to find a clustering that maximizes the entropy difference between clusters. We define this objective as maximizing the minimum inter-cluster distance as defined in Equation~\ref{eq:obj}: 
\begin{equation}
    \max_{C_1, \dots, C_k} (\min_{i \neq j} || x_i - x_j||)
    \label{eq:obj}
\end{equation}
with the constraint that each byte $b_i$ must belong to exactly one cluster: $\bigcup_{j=1}^{k} C_j = X$ and $C_i \cap C_j = \emptyset$ for $i \neq j$. The number of clusters is $1 < k \leq n$. The distance between two clusters $i$ and $j$ is computed using Euclidean distance between the two corresponding features $x_i$ and $x_j$, as shown with $||x_i - x_j||$. 
%
%
%


Lines ~\ref{lin:linkage}--\ref{lin:clustere} in Algorithm~\ref{alg:clustering} shows the clustering step of the \DTT{} algorithm, where it takes the feature list of $X$ and creates a clustering $\mathcal{C}$.
The clustering in Algorithm~\ref{alg:clustering} evaluates clusterings with different numbers of possible clusters and selects the best one with respect to a clustering score. The algorithm starts by assuming each byte as one cluster, which means $n$ clusters. In other words, for single-precision datasets, initially there are $4$ clusters and $8$ clusters for double-precision datasets. In line~\ref{lin:linkage}, the algorithm computes the pairwise distance between every pair of partitions and builds an $n \times n$ matrix $L$. This matrix is called a linkage matrix. Then, the algorithm creates clusterings and their scores for all possible cluster counts, which range from 1 to $n$ (lines~\ref{lin:clusterb}--\ref{lin:clustere}). Finally, the index of the clustering with the maximum score is selected in line~\ref{lin:scores}, where $\operatorname{argmax}$ returns the index of the maximum element in $S$.


The choice of clustering method and score in Algorithm~\ref{alg:clustering} is set experimentally to provide the best clustering with the maximum objective (Equation~\ref{eq:obj}). We find that hierarchical clustering provides the best results.
For the clustering score, \DTT{} supports four well-established internal validation metrics: Silhouette Score~\cite{rousseeuw1987silhouettes}, Davies-Bouldin Score~\cite{davies2009cluster}, Calinski-Harabasz Score~\cite{calinski1974dendrite}, and Gap Statistic~\cite{tibshirani2001estimating}.
Based on the results of this evaluation, we established that for 32-bit floating-point datasets, we primarily rely on the \textit{Davies-Bouldin Score}, and for 64-bit floating-point datasets, we consider using a \textit{Gap Statistic Score} computed across all features.

\subsection{Step III. Packing and Decomposition} 
\label{sec:packing}


Upon completion of the clustering process, the input dataset $D$ will be reorganized according to the clustering information $\mathcal{C}$. This reorganization offers several strategies for arranging bytes adjacent in the transformed dataset based on the selected inter-cluster and intra-cluster ordering.

Intra-cluster byte ordering decides the placement order of bytes within each cluster. We limit this to either group bytes from the same floating value first, i.e. \textit{same-value packing}, or group same bytes across values, i.e. \textit{same-byte packing}. 
We found the two approaches performed similarly with respect to compression ratio. We choose same-byte packing, but it can be defined by the user if needed.


Inter-cluster alignment decides how to place a packed cluster in the transformed dataset. Our findings indicate that the order in which clusters are stored has a negligible impact on the \DTT{} objective. This observation suggests that clusters of bytes can be decomposed, and the offset to the beginning of each byte cluster can be stored.
Decomposition into smaller data arrays with lower entropy offers two advantages: 
\circlednumber{1} Treating each byte cluster as an independent data stream results in a lower average entropy~\cite{cover2006elements} and potentially increases the compression ratio, where the selected method relies on a pure entropy-based compression technique. 
\circlednumber{2} Operating on multiple independent data streams, rather than a single stream, potentially enhances throughput. This approach introduces additional workloads beyond those typically achieved through common data parallelism in compression tools, e.g., blocking. We will discuss how these components are used when combined with a compression tool.  

\section{\DTT{}-based Compression}
\label{sec:compression}

\DTT{} augments generic compression tools with the knowledge of floating-point types to provide a high compression ratio.
Figure~\ref{fig:overall} shows how the transformed dataset is provided as input to a blocked compression tool. This section discusses how data parallelism is used through blocking and how \DTT{} is combined with different compression tools and algorithms. The section also discusses some \DTT{} parameters and selects them to achieve a high compression ratio.

\subsection{Parallel Blocking-based De/Compression}
\label{sec:blocking}
To efficiently integrate \DTT{} into compression methods, this section discusses its combination with blocking to utilize parallel resources. 
Blocking is a common technique for exploiting data parallelism in compression methods. This technique partitions the input data into blocks, where each block can be independently de/compressed in parallel.

To combine blocking with \DTT{}, the order in which clustering and blocking are applied is essential. Blocking can be applied to each cluster of bytes separately, or each block can be clustered independently. We investigate both options and find that applying \DTT{} to each block leads to better throughput with minimal effect on the compression ratio. In other words, the \DTT-based compression first blocks the dataset and then applies transformation per block. This makes sense since applying \DTT{} before chunking does not allow for interleaving data transformation and compression, where threads can be scheduled to use memory and computing resources more efficiently.

Another critical decision for combining \DTT{} with blocking is determining the block size. A larger block size enables a larger window for the compression method, potentially leading to better compression ratios. However, larger blocks also reduce the number of parallel workloads and degrade cache efficiency when they exceed the cache size. We found that the block size should be approximately between L1 and L2 cache sizes to ensure data locality during de/compression. L1 and L2 cache sizes are dedicated to each core in CPUs, hence selected as a limit in this work. The exact size varies from one compression tool to another due to the different memory access patterns of the compression algorithms used. The block size selection also has an impact on load balance, where a larger block size leads to a smaller number of blocks, potentially impacting core utilization. Therefore, we limit the number of blocks, i.e., $\frac{\text{dataset size}}{\text{block size}}$, to be always larger than the number of cores. 
Additionally, all contiguous blocks should be assigned to the same thread to maximize the benefits of prefetching in the cache levels.
%






\begin{table}[!ht]
\centering
\renewcommand{\arraystretch}{1.1}
\setlength{\tabcolsep}{4pt}
\resizebox{\linewidth}{!}{%
\begin{tabular}{l|cc}
\toprule
\textbf{Method} & \textbf{Algorithm/Library} & \textbf{Architecture} \\
\midrule
 \multirow{3}{*}{\makecell[c]{\textbf{Pure}\\\textbf{Foundational}}}  & LZ77 & Single-Thread \\
 & Huffman Coding & Single-Thread \\
  & Delta-based & Single-Thread \\
 \hline
\multirow{5}{*}{\makecell[c]{\textbf{Hybrid}}} & zstd & Multicore CPU \\
 & snappy & Multicore CPU \\
 & lz4 & Multicore CPU \\
 & zlib & Multicore CPU \\
 & bzip & Multicore CPU \\
 & nvCOMP & GPU \\
\bottomrule
\end{tabular}%
}
\caption{Used compression methods/libraries with \DTT{}. }
\label{table:compression_libraries}
\end{table}

\subsection{Compression Methods}
\label{sec:compmethods}
To evaluate the effect of \DTT{} on compression, this section presents the compression methods used in \DTT-based compression. We use two classes of compression methods for this evaluation. We select pure foundational compression algorithms with in-house implementations to demonstrate the effect of \DTT{} on them in isolation. Additionally, we choose generic high-performance compression tools to show the impact of \DTT{} on the performance and compression ratio of hybrid CPU and GPU-based tools, which combine foundational methods with efficient implementations. Table~\ref{table:compression_libraries} provides an overview of the two categories of compression methods studied in this work. This section briefly discusses each category.

\subsubsection{Pure foundational methods}
As shown in Table~\ref{table:compression_libraries}, we select three important compression algorithms to study the effect of \DTT{} in isolation for each of them. The three algorithms are: 
\circlednumber{1} The LZ77 algorithm is a general dictionary-based method, known as a key technique to reduce file sizes by spotting and removing repeated data patterns. LZ77 works by maintaining a moving window over the data being compressed. When it finds a sequence of characters that appears more than once within this window, it replaces the extra copies with a reference to the first instance.
\circlednumber{2} The Huffman coding~\cite{huffman1952method} algorithm is an entropy-based method that assigns a shorter codeword to more frequent symbols. Data transformation does not affect entropy. Therefore, data transformation without decomposing byte clusters does not affect the compression ratio of Huffman codes.
\circlednumber{3} Delta-based methods examine the differences between consecutive floating-point data to exploit redundancy and compress the data. These methods are often used for streaming data when only part of the data is accessible. We use an XOR-based scheme similar to Gorilla~\cite{pelkonen2015gorilla} for this algorithm. 

\subsubsection{Hybrid methods } 
State-of-the-art compression packages often combine different compression algorithms and run them efficiently on parallel CPU cores and vector units. As shown in Table~\ref{table:compression_libraries}, we study zstd~\cite{FacebookZstd}, snappy~\cite{GoogleSnappy}, and lz4~\cite{YannColletLZ4}, zLib~\cite{gailly2004zlib}, and  bzip~\cite{seward1996bzip2} working on multicore CPUs and nvCOMP-LZ4~\cite{nvcomp} for GPUs as representatives of efficient combinations of dictionary and entropy coders. The selected compression tools are a combination of run-length encoding (RLE) and a dictionary-based coding and/or an entropy coder such as Huffman coding~\cite{huffman1952method} and benefit from common data transformation. 

These selected compression tools also have different performance throughput and compression ratios on multicore and GPUs. We provide a short summary of each tool: 
\circlednumber{1} zstd uses LZ77 algorithm followed by Finite State Entropy (FSE)~\cite{FacebookZstd} and Huffman Coding which are entropy coders. FSE is an entropy coder that encodes and decodes processes based on transitions between a finite number of states. These states hold information about the probability distribution of the symbols encountered so far.
zstd provides different levels of compression to ensure a compromise between compression ratio and throughput.  
\circlednumber{2} snappy is a compression method based on dictionary-coding, similar to LZ77. snappy focuses on throughput by reducing the window search for finding patterns and using a simple greedy approach.
\circlednumber{3} lz4 is also following a similar scheme to snappy, with more emphasis on throughput.  %
\circlednumber{4} zlib implements the DEFLATE algorithm~\cite{oswal2016deflate} which is a combination of LZ77 and Huffman coding. 
\circlednumber{5} bzip combines RLE, BWT~\cite{manzini2001analysis}, move-to-front transform~\cite{arnavut2000move}, and Huffman coding which degrades its throughput.
\circlednumber{6} nvCOMP is a vendor library provided by NVIDIA and is commonly used to compress data using a hybrid of foundational compression methods. nvCOMP has implementations of several compression algorithms. We use lz4 variant of nvCOMP for fair comparison with its CPU variant where needed.


\subsection{Implementation}
This section discusses some implementation details to reduce clustering overhead using compression modes and to encode the \DTT{} transformation details in the compressed data.

\subsubsection{Compression Modes}
\label{sec:modes}
To avoid the cost of clustering during compression, we provide two modes for \DTT-based compression: dynamic and static modes. The dynamic mode relies on clustering for each dataset in the compression phase. For when the size of the data is big, the cost of clustering amortizes over the compression time. We apply clustering on a portion of the dataset. We find that 30\% of the dataset is sufficient to capture the properties of the dataset. The static mode relies on clustering information based on clustering per application. This clustering information is collected through extensive profiling in our datasets from several domains. To calculate clustering in static mode, datasets are first categorized by application and their type width. Then, a sample of each dataset is merged to make a compound dataset. The clustering is then applied to the compound dataset and used for all datasets of the category. These two modes provide a trade-off between compression throughput and compression ratio in \DTT-based compression.

\subsubsection{Compression format}
The \DTT-based compression encodes all metadata in the compressed data with small overhead. The metadata encoded in the compressed dataset is the clustering information, selected compression methods, and block size. 
The compression tool is encoded as an integer ID. 
The compressed blocks and byte clusters are stored with their offset, e.g., as a 2D array with variable length dimensions, so accessing to each compressed part in a block is easily accessible, essential for parallel decompression. So during decompression, first the compression tool is encoded, and then block offsets and byte cluster offsets are used to locate compressed data and then decompress them in parallel.

\begin{table}[h!]
\centering
\renewcommand{\arraystretch}{1.1}
\setlength{\tabcolsep}{4pt}
\resizebox{\linewidth}{!}{%
\begin{tabular}{lccccc}
\toprule
\textbf{ID} & \textbf{Application} & \textbf{Type (S/D)} & \textbf{Size (bytes)} & \textbf{Entropy} & \textbf{Domain} \\
\midrule
\rowcolor{lightgray} D1 & astro-mhd & D & 548M & 0.97 & HPC \\
D2 & rsim & S & 94M & 18.50 & HPC \\
\rowcolor{lightgray} D3 & turbulence & S & 67M & 23.73 & HPC \\
D4 & wave & S & 537M & 25.27 & HPC \\
\rowcolor{lightgray} D5 & num-brain & D & 142M & 23.97 & HPC \\
D6 & num-control & D & 160M & 24.14 & HPC \\
\rowcolor{lightgray} D7 & astro-pt & D & 671M & 26.32 & HPC \\
D8 & msg-bt & D & 266M & 23.67 & HPC \\
\rowcolor{lightgray} D9 & citytemp & S & 12M & 9.43 & TS \\
D10 & wesad-chest & D & 272M & 13.85 & TS \\
\rowcolor{lightgray} D11 & solar-wind & S & 424M & 14.06 & TS \\
D12 & hdr-night & S & 537M & 9.03 & OBS \\
\rowcolor{lightgray} D13 & hdr-palermo & S & 843M & 9.34 & OBS \\
D14 & hst-wfc3-ir & S & 24M & 15.04 & OBS \\
\rowcolor{lightgray} D15 & hst-wfc3-uvis & S & 109M & 15.61 & OBS \\
D16 & acs-wht & S & 225M & 20.13 & OBS \\
\rowcolor{lightgray} D17 & spitzer-irac & S & 165M & 20.54 & OBS \\

D18 & jws-mirimage & S & 169M & 23.16 & OBS \\
\rowcolor{lightgray} D19 & tpcxBB-store & D & 790M & 16.73 & DB \\
D20 & tpcxBB-web & D & 987M & 17.64 & DB \\
\rowcolor{lightgray} D21 & tpcH-order & D & 120M & 23.40 & DB \\
D22 & tpcDS-catalog & S & 173M & 17.34 & DB \\
\rowcolor{lightgray} D23 & tpcDS-store & S & 277M & 15.17 & DB \\
D24 & tpcDS-web & S & 86M & 17.33 & DB \\
\rowcolor{lightgray} D25 & Llama~\cite{touvron2023llama} & H & 48G & 4.88 & ML \\
\bottomrule
\end{tabular}%
}
\caption{Combined Datasets from HPC, TS, OBS, and DB Domains with Application and ID Columns (*S/D denotes single-/double-precision).}
\label{tab:combined_application_with_id}
\end{table}

\begin{table*}[!ht]
\centering
\resizebox{\linewidth}{!}{%
\begin{tabular}{l l c c  c c  c c  c c  c c  c c}
\toprule
\multirow{2}{*}{\textbf{ID}} & \multirow{2}{*}{\textbf{Dataset}} 
  & \multicolumn{2}{c}{\textbf{zstd}} 
  & \multicolumn{2}{c}{\textbf{snappy}} 
  & \multicolumn{2}{c}{\textbf{lz4}} 
  & \multicolumn{2}{c}{\textbf{zlib}} 
  & \multicolumn{2}{c}{\textbf{bzip2}} 
  & \multicolumn{2}{c}{\textbf{nvCOMP}} \\
\cmidrule(lr){3-4} \cmidrule(lr){5-6} \cmidrule(lr){7-8} \cmidrule(lr){9-10} \cmidrule(lr){11-12} \cmidrule(lr){13-14}
 &  & \textbf{TDT} & \textbf{Standard} 
        & \textbf{TDT} & \textbf{Standard} 
        & \textbf{TDT} & \textbf{Standard} 
        & \textbf{TDT} & \textbf{Standard} 
        & \textbf{TDT} & \textbf{Standard} 
        & \textbf{TDT} & \textbf{Standard} \\
\midrule
\rowcolor{lightgray}
D1  & astro\_mhd       & 33.146 & 29.871 & 12.934 & 12.250 & 28.540 & 25.216 & 30.711 & 27.753 & 28.621 & 25.216 & 18.910 & 22.670 \\
D2  & rsim             & 1.805  & 1.483  & 1.638  & 1.297  & 1.745  & 1.358  & 1.745  & 1.451  & 1.782  & 1.440  & 1.580  & 1.310  \\
\rowcolor{lightgray}
D3  & turbulence       & 1.252  & 1.093  & 1.199  & 1.000  & 1.209  & 0.996  & 1.264  & 1.092  & 1.285  & 1.063  & 1.130  & 1.000  \\
D4  & wave             & 1.794  & 1.253  & 1.453  & 1.000  & 1.358  & 1.077  & 1.676  & 1.154  & 1.421  & 1.164  & 1.280  & 1.030  \\
\rowcolor{lightgray}
D5  & num\_brain       & 1.265  & 1.065  & 1.209  & 1.000  & 1.234  & 0.997  & 1.262  & 1.064  & 1.241  & 0.997  & 1.200  & 1.000  \\
D6  & num\_control     & 1.146  & 1.058  & 1.100  & 1.010  & 1.107  & 1.013  & 1.154  & 1.057  & 1.154  & 1.030  & 1.020  & 1.010  \\
\rowcolor{lightgray}
D7  & astro\_pt        & 1.366  & 1.047  & 1.262  & 1.000  & 1.320  & 0.999  & 1.330  & 1.048  & 1.331  & 0.999  & 1.260  & 1.000  \\
D8  & msg\_bt          & 1.383  & 1.126  & 1.277  & 1.052  & 1.318  & 1.065  & 1.345  & 1.131  & 1.324  & 1.065  & 1.270  & 1.060  \\
\rowcolor{lightgray}
D9  & citytemp         & 2.987  & 2.665  & 2.023  & 1.617  & 2.013  & 1.974  & 3.117  & 2.530  & 3.589  & 3.882  & 1.600  & 1.370  \\
D10 & wesad\_chest     & 3.305  & 3.349  & 2.257  & 2.351  & 2.785  & 2.953  & 3.563  & 3.344  & 2.810  & 2.953  & 1.810  & 2.130  \\
\rowcolor{lightgray}
D11 & solar\_wind      & 1.708  & 1.440  & 1.375  & 1.111  & 1.491  & 1.343  & 1.811  & 1.582  & 1.998  & 2.021  & 1.290  & 1.170  \\
D12 & hdr\_night       & 3.647  & 2.842  & 2.855  & 1.832  & 3.176  & 2.178  & 3.679  & 2.716  & 3.822  & 3.977  & 2.700  & 1.400  \\
\rowcolor{lightgray}
D13 & hdr\_palermo     & 4.329  & 3.451  & 3.304  & 1.947  & 4.000  & 2.391  & 4.437  & 3.196  & 5.647  & 5.598  & 3.250  & 1.420  \\
D14 & hst\_wfc3\_ir    & 2.011  & 1.773  & 1.867  & 1.478  & 1.962  & 1.526  & 2.014  & 1.761  & 2.105  & 1.743  &       &        \\
\rowcolor{lightgray}
D15 & hst\_wfc3\_uvis & 1.935  & 1.756  & 1.749  & 1.555  & 1.839  & 1.610  & 1.935  & 1.749  & 1.955  & 1.702  & 1.710  & 1.540  \\
D16 & acs\_wht         & 1.580  & 1.369  & 1.437  & 1.142  & 1.478  & 1.166  & 1.586  & 1.362  & 1.612  & 1.429  & 1.410  & 1.160  \\
\rowcolor{lightgray}
D17 & spitzer\_irac    & 1.174  & 1.076  & 1.104  & 1.000  & 1.110  & 0.996  & 1.178  & 1.074  & 1.180  & 1.044  & 1.050  & 1.000  \\
D18 & jw\_mirimage     & 1.313  & 1.148  & 1.287  & 1.000  & 1.299  & 0.999  & 1.322  & 1.145  & 1.318  & 1.150  & 1.270  & 1.000  \\
\rowcolor{lightgray}
D19 & tpcxbb\_store    & 2.561  & 2.333  & 1.878  & 1.644  & 2.204  & 2.028  & 2.625  & 2.470  & 2.170  & 3.130  & 1.530  & 1.730  \\
D20 & tpcxbb\_web      & 2.485  & 2.224  & 1.830  & 1.611  & 2.167  & 1.986  & 2.548  & 2.382  & 2.110  & 2.960  & 1.510  & 1.690  \\
\rowcolor{lightgray}
D21 & tpch\_order      & 2.091  & 1.836  & 1.502  & 1.359  & 1.659  & 1.598  & 2.113  & 1.864  & 1.662  & 1.598  & 1.220  & 1.500  \\
D22 & tpcds\_catalog   & 1.460  & 1.238  & 1.156  & 1.057  & 1.229  & 1.113  & 1.440  & 1.301  & 1.474  & 1.520  & 1.110  & 1.060  \\
\rowcolor{lightgray}
D23 & tpcds\_store     & 1.554  & 1.353  & 1.248  & 1.155  & 1.370  & 1.253  & 1.610  & 1.450  & 1.800  & 1.713  & 1.130  & 1.110  \\
D24 & tpcds\_web       & 1.461  & 1.238  & 1.156  & 1.057  & 1.230  & 1.113  & 1.441  & 1.301  & 1.475  & 1.520  & 1.110  & 1.060  \\
\rowcolor{lightgray}
D25 & LLama            & 1.912  & 1.788  & 1.426  & 1.071  & 1.399  & 1.429  & 1.970  & 1.743  & 1.910  & 2.090  &       &        \\
\bottomrule
\end{tabular}%
}
\caption{Compression ratio comparison (TDT vs.\ standard) }
\label{tab:CompressionRatio}
\end{table*}

\section{Experimental Results}
This section evaluates the efficiency of \DTT{}-based compression using a wide spectrum of datasets and compression tools. 
Particularly, this section answers the following fundamental questions: \circlednumber{1} What is the effect of \DTT{} on compression ratios of different foundational and hybrid compression methods across domains? \circlednumber{2} How does \DTT{} affect throughput of compression and decompression when combined with other methods in CPU and GPUs? \circlednumber{3} How are \DTT{} critical components, such as clustering method, feature selection, and blocking selected?  




\subsection{Setup}

\subsubsection{Dataset}
The list of datasets used for \DTT{} is provided in Table~\ref{tab:combined_application_with_id}. 
We apply \DTT{} to a list of datasets from different applications obtained from FCbench~\cite{chen2024fcbench}. The selected datasets are from HPC~\cite{kissmann2016colliding,thoman2020rtx}, TS~\cite{schmidt2018introducing}, OBS~\cite{rsaspitzerdocs}, and DB~\cite{tpc_tpcds}. While selected FCBench datasets have both double precision (D or float64) and single precision (S or float32), it does not have half-precision. We use a trained language model, Llama 7B~\cite{touvron2023llama}, and extract its weights and attention matrices with half precision (H) and store them all as one big dataset. 

\subsubsection{Platform and Measurements}
We evaluate our method on both CPU and GPU architectures. We select a consumer-grade CPU and GPU for this purpose. The CPU is \textit{Core-i9} with 16 cores, 760 KB, 24MB, and 30MB L1/L2/L3 cache sizes. We use NVIDIA GeForce RTX 4090 with Ampere architecture. The GPU has 3584 Cuda cores with L1/L2 cache sizes of 128 KB per SM and 3 MB, respectively.  
A \texttt{C++} benchmark is created to drive all compression tools and \DTT{} for fair comparison. The benchmark is built using \texttt{CMake} and GCC compiler version 11.4.0. All experiments are performed on Ubuntu 22.04.3 LTS. We also implemented \DTT{} in Python to verify the compression ratio with Python libraries of the compression tools. 
For parallel implementations of \DTT{}, \texttt{OpenMP} is selected for CPUs and \texttt{Cuda} is used for GPU where needed. A close thread binding is selected to ensure each thread is mapped to a physical core. 






To assess compression performance, we use three key metrics: \textit{compression ratio (CR)}, \textit{compression throughput (CT)}, and \textit{decompression throughput (DT)}. These are defined as $\text{CR} = \frac{\text{Original Size}}{\text{Compressed Size}}$, $\text{CT} = \frac{\text{Original Size}}{\text{Compression Time}}$, and $\text{DT} = \frac{\text{Original Size}}{\text{Decompression Time}}$, consistent with prior work~\cite{chen2024fcbench}.
In addition to these absolute metrics, we introduce \textit{compression ratio improvement (CRI)} as a relative measure. CRI quantifies the effect of our approach's (specifically, \DTT{}'s) compression ratio compared to a baseline, defined as $\text{CRI} = \frac{\text{\DTT{} CR}}{\text{Baseline CR}}$. We similarly define a relative throughput improvement metric as compression throughput (CTI) and decompression throughput (DTI).

\subsubsection{Compression tools} \DTT{} is combined with compression tools provided in Table~\ref{table:compression_libraries}. 
Foundational methods, except for LZ77, are from an in-house proof-of-concept, only for compression ratio analysis. Due to the inefficiency of these implementations, they are excluded from throughput evaluation. For LZ77, we use the latest version of fastlz\footnote{\url{https://github.com/ariya/FastLZ}}, which an efficient single-threaded implementation of the LZ77 algorithm.
We use bzip v1.0.8, lz4 v1.9.4, nvComp v4.1.1.1, and the latest release of snappy, zlib, and zstd for comparison. Public repositories of bzip\footnote{\url{https://gitlab.com/bzip2/bzip2}}, lz4\footnote{\url{https://github.com/lz4/lz4}}, snappy\footnote{\url{https://github.com/google/snappy}}, zlib\footnote{\url{https://github.com/madler/zlib}}, zstd\footnote{\url{https://github.com/facebook/zstd}} are used to obtain the latest version or release of compression tools. NVComp is a proprietary tool, and we used the CUDA toolkit version 11 for comparison. 
All compression tools are used with their default configurations. All numbers are reported \DTT{} static mode unless otherwise stated. 


\begin{figure}[h!]
  \centering
 \includegraphics[width=\linewidth, height=4.5cm]{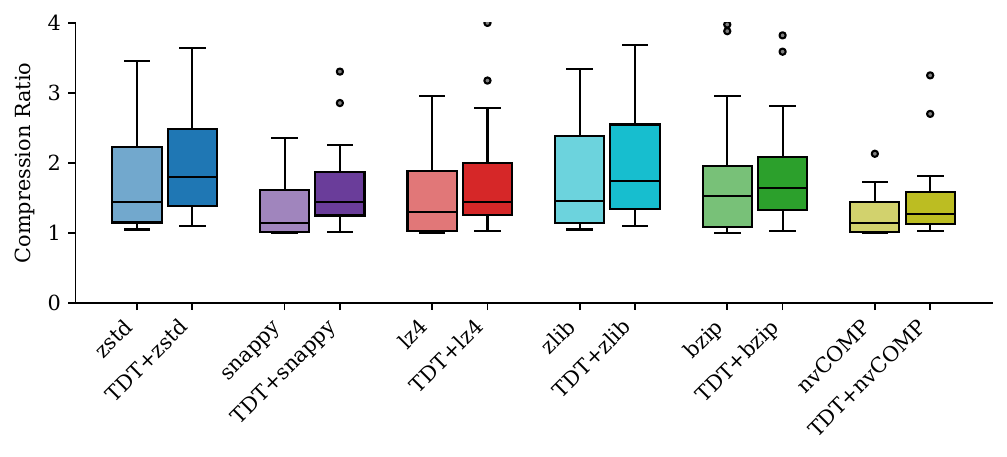}
  \caption{\DTT{} improves CR across datasets. Some outlier points are cropped for better illustration.  }
  \label{fig:CROverall}
\end{figure}

  \begin{figure*}[!t] 
    \centering
    \includegraphics[width=0.3\linewidth, height=2cm]{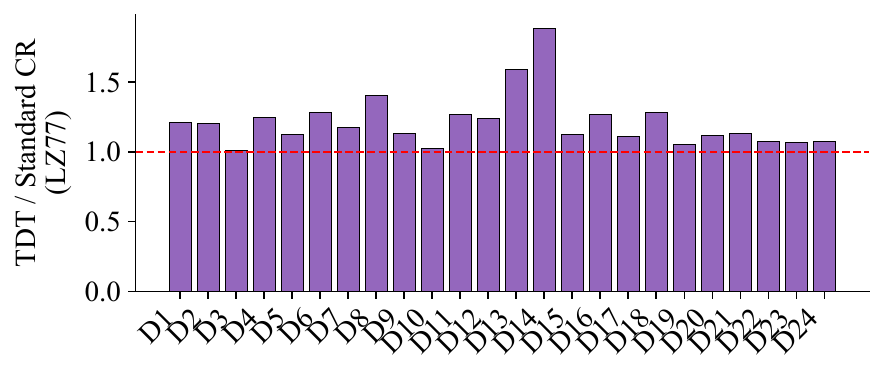}
    \includegraphics[width=0.3\linewidth, height=2cm]{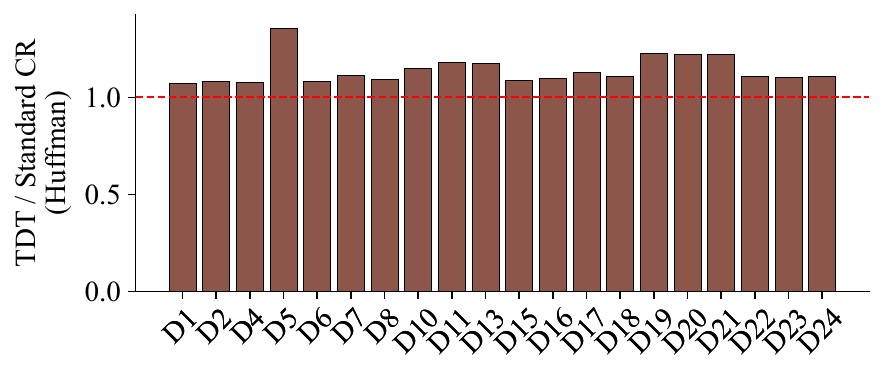}
    \includegraphics[width=0.3\linewidth, height=2cm]{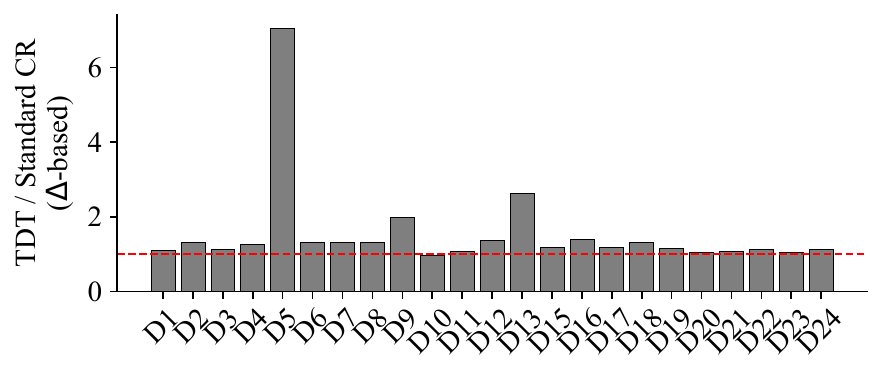}

  \caption{\DTT{} improves compression ratio of foundational methods (LZ77, Huffman, Delta-based)} 

  \label{fig:CRfoundational}
\end{figure*}

\begin{figure*}[!t] 
    \centering
    \includegraphics[width=0.3\linewidth, height=2cm]{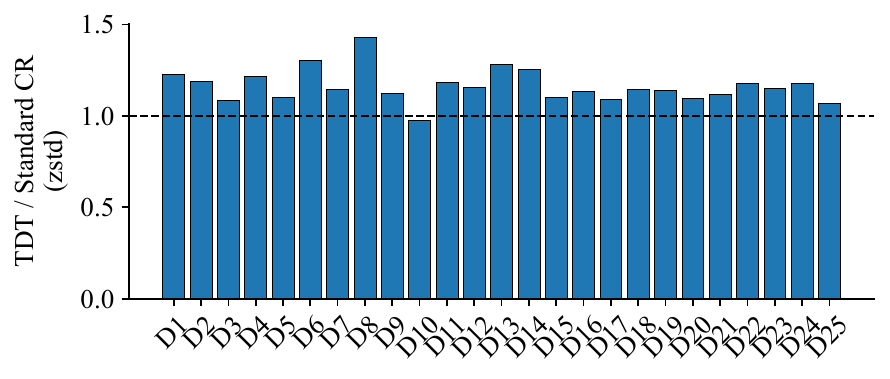}
    \includegraphics[width=0.3\linewidth, height=2cm]{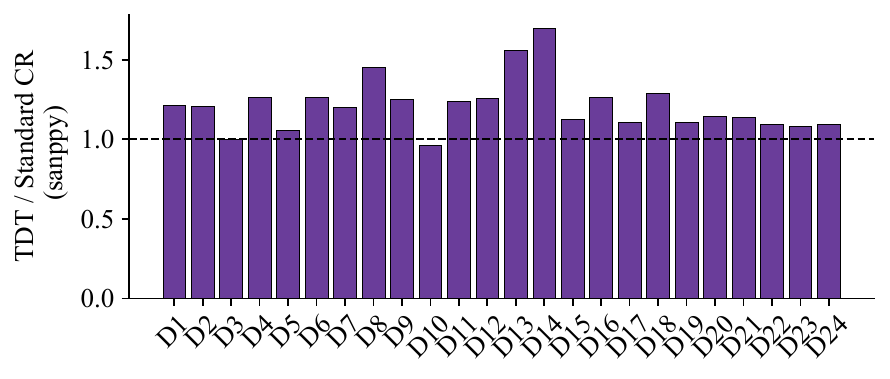}
    \includegraphics[width=0.3\linewidth, height=2cm]{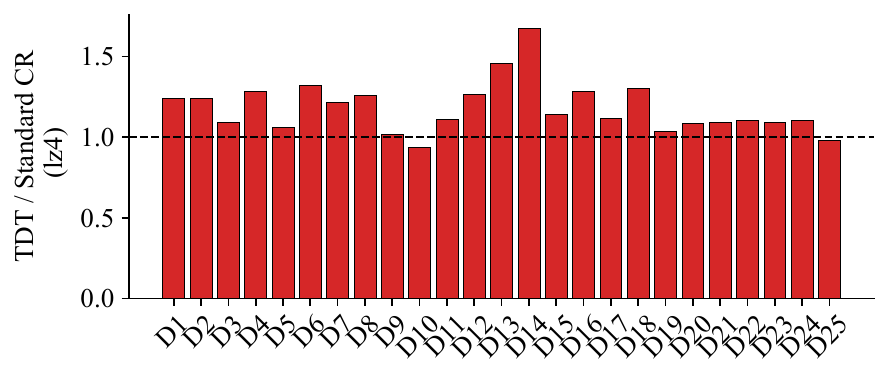}
    \includegraphics[width=0.3\linewidth, height=2cm]{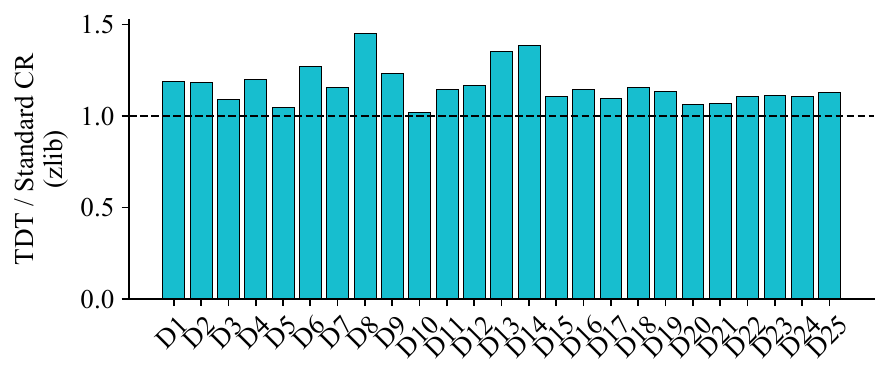}
    \includegraphics[width=0.3\linewidth, height=2cm]{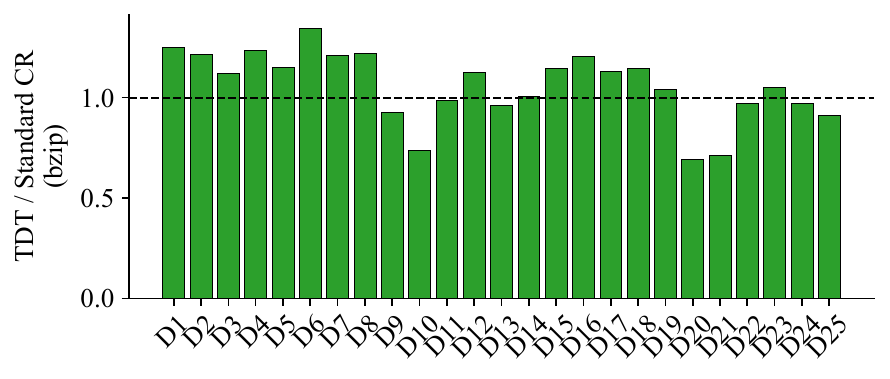}
     \includegraphics[width=0.3\linewidth, height=2cm]{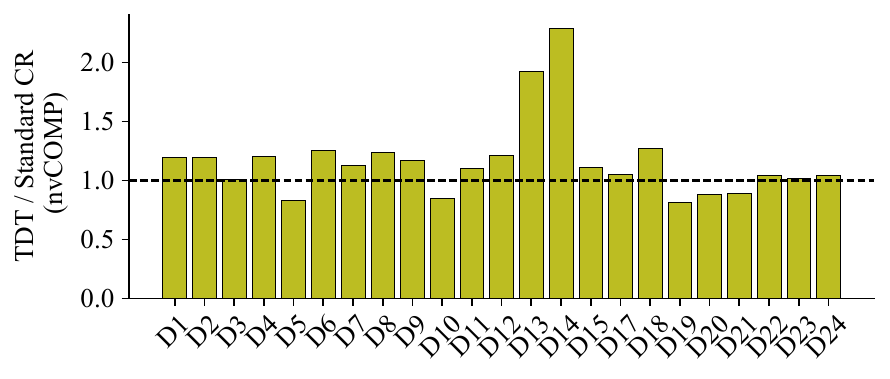}
    \caption{\DTT{} Compression Ratio vs. standard Compression ratio Across Datasets for zstd, snappy, lz4, zlib, bzip, and nvCOMP.}
    \label{fig:Ratio}
\end{figure*}

\subsection{Compression Ratio Evaluation}
This section presents how \DTT-based compression performs across datasets and tools. 

\subsubsection{Overall comparison} 
Figure~\ref{fig:CROverall} shows an overall view of the distribution of compression ratios for each hybrid compression tool. The x-axis shows the compression tools, and the y-axis shows the compression ratio across all datasets in Table~\ref{tab:combined_application_with_id} for the tool. As shown, \DTT{} improves the compression ratio of all tools across datasets. On average, \DTT{} provides between 1.04--1.20$\times$ geometric mean CRI across tools. Table~\ref{tab:CompressionRatio} shows the exact compression ratio for each dataset and compression tool. In the rest of this section, we analyze these data from different perspectives.  


\subsubsection{Effect of \DTT{} on foundational methods} Figure~\ref{fig:CRfoundational} shows the effect of \DTT{} improves compression ratios on different foundational methods discussed in Table~\ref{table:compression_libraries}. As shown in Figure~\ref{fig:CRfoundational}, the compression ratio for all foundational methods increases. The CRI for \DTT{} for LZ77, Huffman, and Delta-based methods are 1.20, 1.14, and 1.35$\times$, respectively. \DTT{} benefits LZ77 more than Huffman as it takes advantage of both reordering and decomposition.   



\subsubsection{Effect of \DTT{} on hybrid methods} Figure~\ref{fig:Ratio} illustrates how \DTT{} improves compression ratios of datasets across compression tools. While there is consistent improvement across all datasets and tools, the effect of the transformation varies across tools. For bzip, the CRI is lower than for other tools. This is attributed to the internal BWT data transformation used in bzip, which reduces the effect of \DTT. However, other compression methods exhibit similar behavior due to the combination of compression categories, i.e., dictionary-based and entropy coders.

Datasets exhibit different properties that lead to different CRI when \DTT{} is used. For example, datasets 4 and 6 consistently show the highest CR for all five compression methods. We find these datasets have the highest entropy, larger than $23$, which restricts existing tools from compressing them. After applying \DTT, the opportunities for compression are enabled and utilized, leading to higher CRI.

\subsubsection{Entropy Analysis}
\begin{figure}
    \centering
    \begin{subfigure}[b]{\linewidth}
        \includegraphics[width=0.9\linewidth]{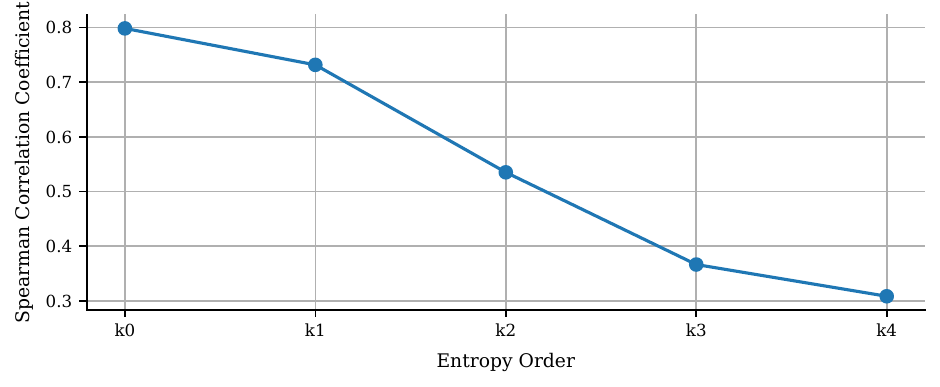}
        \caption{Spearman correlation coefficient between the zstd compressed size after \DTT{} with different entropy orders of datasets}
        \label{fig:spearman}
    \end{subfigure}
    \begin{subfigure}[b]{\linewidth}
        \includegraphics[width=0.9\linewidth]{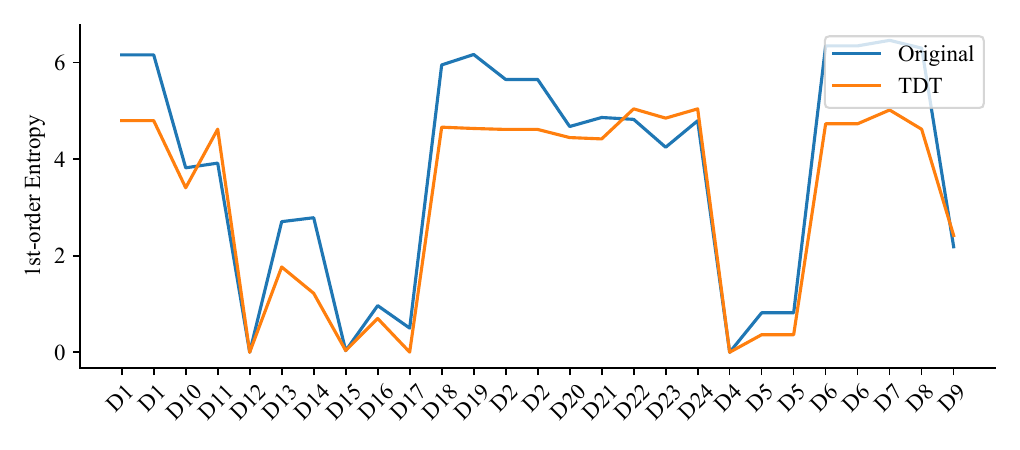}
        \caption{Actual 1st-order entropy  for all datasets for zstd}
        \label{fig:secondOrderEntropy}
    \end{subfigure}
    \caption{Entropy Analysis}
    \label{fig:entropyanalysis}
\end{figure}

The main cause for improving CR is that \DTT{} improves the $k^{th}$ order entropy. We compute the $k^{th}$ order entropy for $k=0, 1, 2, 3, 4$ and observe that while zero-order entropy is the same across datasets after \DTT{}, the effect of \DTT{} is noticeable in higher orders. Figure~\ref{fig:spearman} shows the Spearman Correlation Coefficient between the compressed size of datasets using \DTT{} and zstd across different entropy orders. Coefficients closer to one indicate a better correlation. As shown, there is a strong correlation for zero and first-order entropy and a moderate correlation for second-order entropy. Therefore, we use first and second-order entropy information to show the effect of \DTT{}. Figure~\ref{fig:secondOrderEntropy} shows that the first-order entropy of the datasets after transforming with \DTT{} is almost always lower than the original dataset's first-order entropy. This explains why \DTT{} improves CR across datasets.

\begin{figure}[h!]
  \centering
  \includegraphics[width=0.9\linewidth]{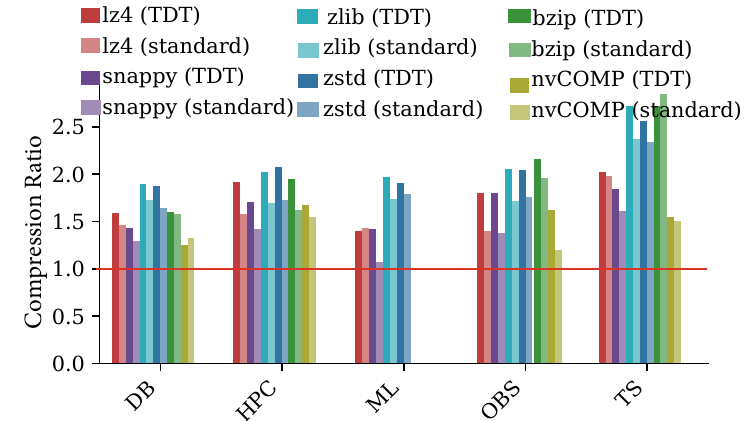}
  \caption{\DTT{} compression ratio across applications. }
  \label{fig:CRapplication}
\end{figure}

\subsubsection{Effect of \DTT{} on application categories} As depicted in Figure~\ref{fig:CRapplication}, \DTT{} performs differently across various application categories. 
Datasets from different applications provide a different range of entropy values. As shown in Table~\ref{tab:combined_application_with_id}, time-series datasets have the lowest entropy values, while DB, OBS, and HPC have relatively higher entropy values than time-series datasets.
 Certain compression tools also perform better with specific applications, such as zlib, which performs better when applied to DB datasets.


\begin{figure}[htbp]
    \centering
    \includegraphics[width=1\linewidth]{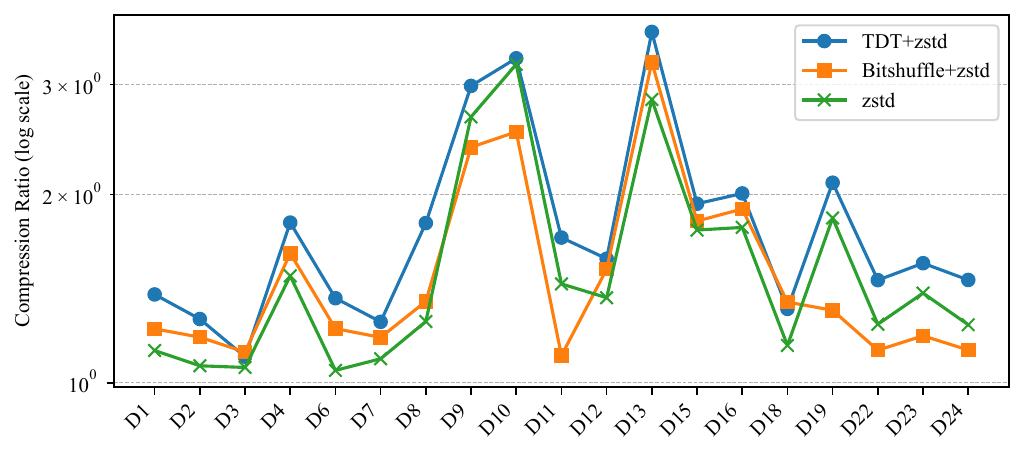}
    \caption{bitshuffle compression ratio across datasets}
    \label{fig:bitshuffle}
\end{figure}

\subsubsection{Comparing with prior floating point compressions} 
\begin{figure}
    \centering
    \includegraphics[width=0.8\linewidth]{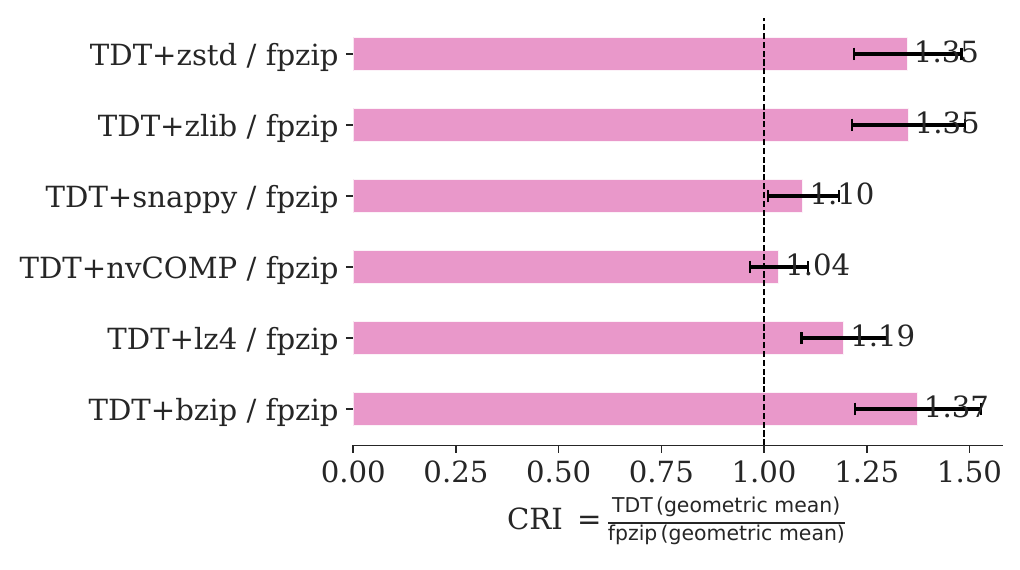}
    \caption{comparing CR with fpzip}
    \label{fig:fpzip}
\end{figure}

Figure~\ref{fig:bitshuffle} and \ref{fig:fpzip} compare \DTT{} with existing data transformations for floating-point data, specifically bitshuffle~\cite{masui2015compression} and a floating point compression tool fpzip. bitshuffle and fpzip perform the best comparing with other tools such as pFPC~\cite{burtscher2008fpc} and SPDP~\cite{claggett2018spdp}, hence selected for comparison. We use existing implementations of bitshuffle\footnote{\url{https://github.com/h5py/h5py}}. For a fair comparison, the CR values in Figure~\ref{fig:bitshuffle} are reported when both data transformations are used with the zstd. As shown, \DTT{} performs better than the bitshuffle data transformations across all applications. The GMean CRI of \DTT{} over bitshuffle is 1.24$\times$. 

Figure~\ref{fig:fpzip} compares GMean CR for different variants of \DTT-based with fpzip. Although fpzip is tuned specifically for floating‐point streams, combining \DTT{} with a generic compressor still yields consistent wins. In datasets with nonuniform value distributions or repeated patterns, \DTT’s block‐rearrangement step produces more homogeneous chunks, allowing the compressor to capture redundancy that fpzip’s predictive model alone cannot. In addition to a higher GMean, \DTT{} variants provide a better CR over fpzip in over 99\% of datasets.



\begin{figure}[h!]
  \centering
  \includegraphics[width=1\linewidth, height=4cm]{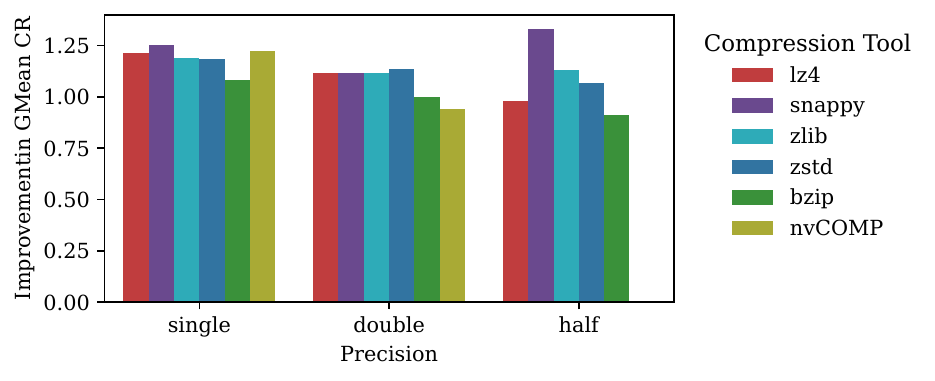}
  \caption{\DTT{} compression ratio across different precisions }
  \label{fig:CRprecision}
\end{figure}

\subsubsection{Floating-point type width effect} Figure~\ref{fig:CRprecision} shows the effect of \DTT{} on floating-point datasets with double, single, and half precision. The GMean CRI for double, single, and half-precision is 1.075, 1.19, and 1.08, respectively, across hybrid methods and all datasets. As shown, this trend is consistent across compression tools.  Additionally, for half-precision datasets, the clustering will have only one option, which is two clusters, where each has one byte. While simple, \DTT{} improves the compression ratio in half-precision datasets across tools.



\begin{figure}
    \centering
   \includegraphics[width=1\linewidth, height=4cm]{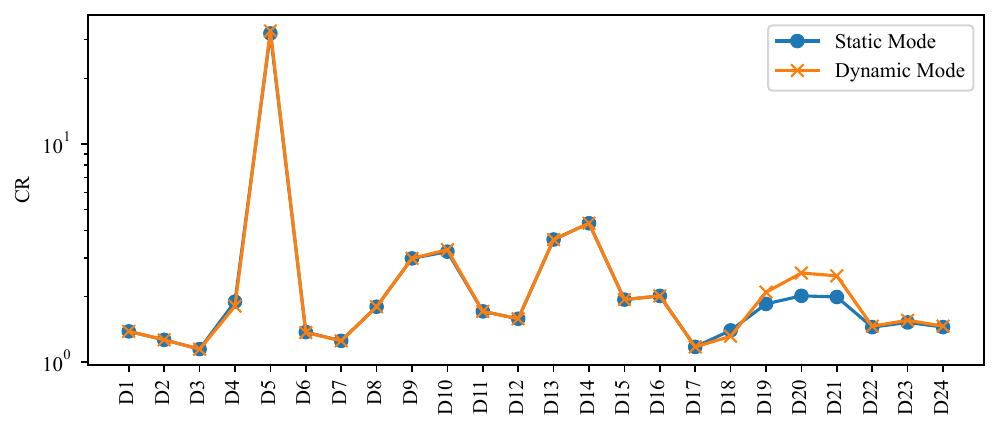}
    \caption{Effect of Dynamic vs. Statistical Mode on Compression Ratio (log scale) using zstd.}
    \label{fig:compModes}
\end{figure}

\subsubsection{Compression modes} As discussed in Section~\ref{sec:modes}, \DTT-based compression has dynamic and static modes.  The compression ratios of the two modes when the compression tool is zstd are shown in Figure~\ref{fig:compModes}. As shown~\ref{fig:compModes}, the compression ratio across datasets follows a similar trend, indicating the efficiency of the static mode, causing zero overhead. However, dynamic mode has its use case when the data pattern is unknown.


\subsection{Throughput Evaluation}
This section discusses the effect of \DTT{} on the compression and decompression throughput when combined with hybrid compression tools shown in Table~\ref{table:compression_libraries}. 

\begin{figure}[htbp]
    \centering
    \begin{subfigure}[b]{1\linewidth}
        \includegraphics[width=1\linewidth]{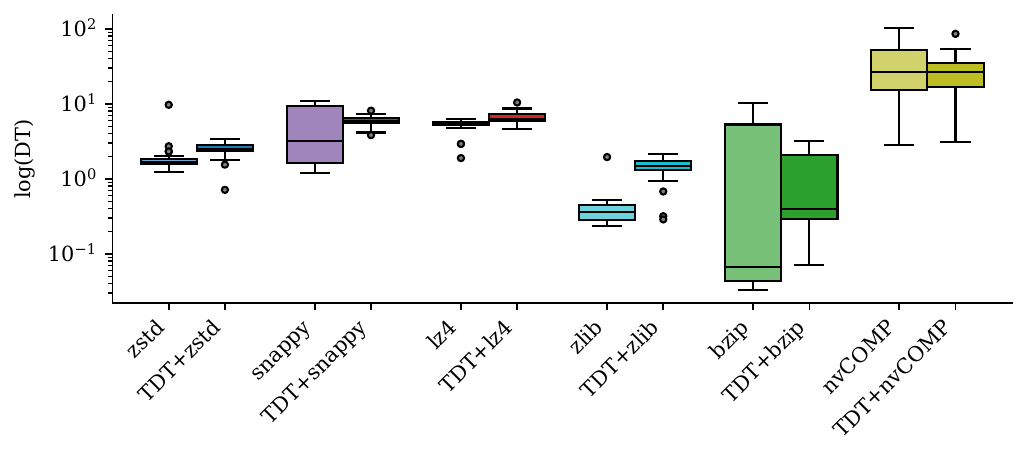}
        \caption{Decompression }
        \label{fig:decompression_Throughput}
    \end{subfigure}
        \begin{subfigure}[b]{1\linewidth}
        \includegraphics[width=1\linewidth]{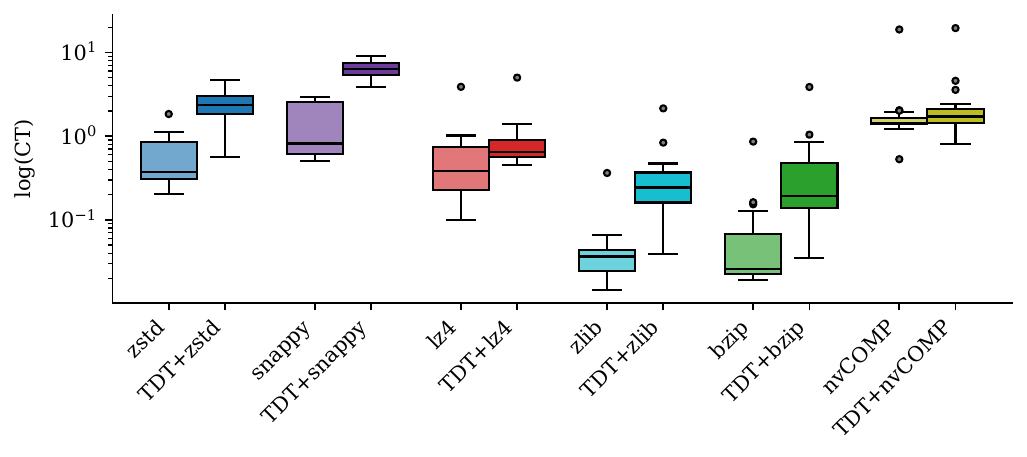}
        \caption{Compression }
        \label{fig:compression_Throughput}
    \end{subfigure}
    
    \caption{\DTT{} throughput in GB/s (some outlier points are cropped for better illustration)}
    \label{fig:Throughput}
\end{figure}

\subsubsection{Overall} Figure~\ref{fig:Throughput} provides an overview of the effect of \DTT{} on the compression and decompression throughput when combined with different compression tools. As shown, when combined with \DTT{}, the throughput of both compression and decompression is improved on average in all CPU and GPU methods, with GMean CTI ranging from 0.97$\times$ to 6.79$\times$ and GMean DTI ranging from 0.59$\times$ to 4.91$\times$ when using \DTT{} compared to the standard baseline.
The improvement is due to the packing and decomposition step of \DTT{}, which creates more independent workloads for each block, enhancing parallelism and load balancing.

As shown in Figure~\ref{fig:Throughput}, the compression throughput is lower than the decompression throughput in all tools. The GMean compression and decompression throughput improvement are 0.96$\times$ and 3.09$\times$, respectively, indicating a more significant effect of \DTT{} on the compression throughput. One reason for the difference is that the compression phase in dictionary-based compression, which is common in all hybrid methods except for bzip, takes more computation, and packing enables more parallelism and better use of resources. The de/compression throughput follows a different trend in bzip under \DTT{} because the method relies on memory-bound operations, and packing and parallelism do not help as much as with other compression techniques.

\subsubsection{CPU vs GPU efficiency} 
The throughput of GPU methods is better than CPU methods thanks to their extensive computing cores. \DTT{} improves the GPU compression and decompression throughput by a GMean of 1.93$\times$ and 24.49$\times$, respectively. This indicates \DTT{} effectiveness on GPUs as well as CPUs. 



\subsection{\DTT{} Empirical Design Choices}
\DTT{} and its compression pipeline are based on some parameters that are selected empirically. This section discusses the design choices in \DTT{} and the \DTT-based compression pipeline which are feature selection (Section~\ref{sec:feature}), clustering parameters (Section~\ref{sec:clustering}), packing (Section~\ref{sec:packing}), and block size (Section~\ref{sec:blocking}).

    


\begin{figure}[htbp]
    \centering
    \begin{subfigure}[b]{0.48\linewidth}
        \includegraphics[width=\linewidth]{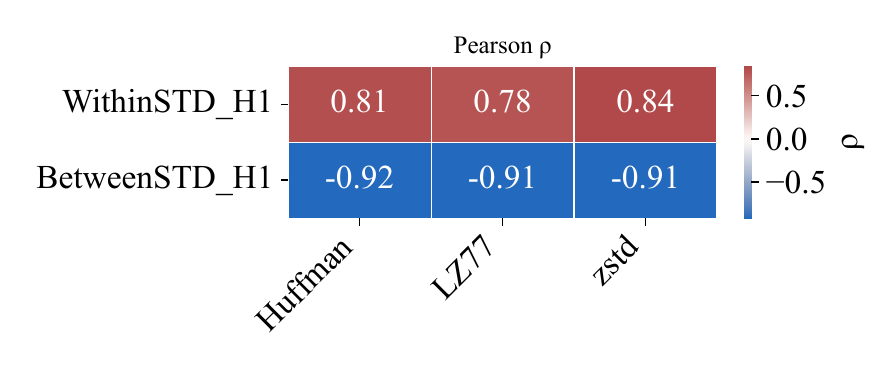}
        \caption{Objective vs. 1st order entropy}
        \label{fig:correlation_H2}
    \end{subfigure}
    \hfill
    \begin{subfigure}[b]{0.51\linewidth}
        \includegraphics[width=\linewidth]{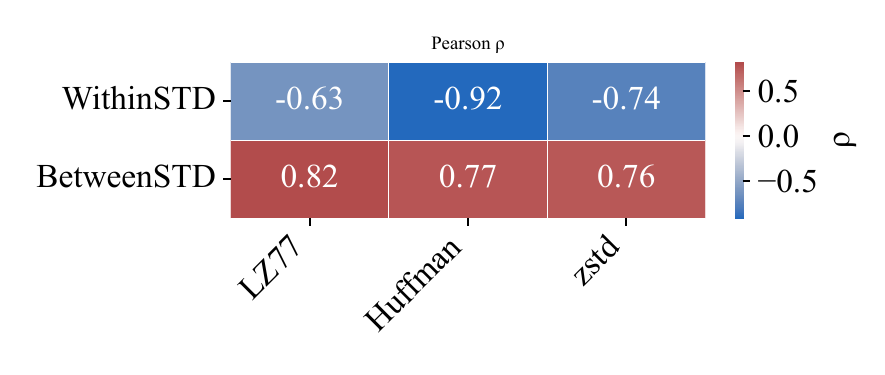}
        \caption{Objective vs. CR}
        \label{fig:STD_H}
    \end{subfigure}
    \caption{The choice of the clustering objective}
    \label{fig:std-entropy}
\end{figure}

\subsubsection{Clustering objective} To justify the effectiveness of the clustering objective in Equation~\ref{eq:obj}, we provide two correlation plots in Figure~\ref{fig:std-entropy}. Figure~\ref{fig:correlation_H2} shows a high correlation between the objective and reducing the first-order entropy deviation intra-cluster and increasing the first-order entropy deviation between clusters. This complies with the objective of \DTT{} to improve the first-order entropy. Also, even though Equation~\ref{eq:obj} focused on between clusters, the deviation within clusters is minimized. This entropy minimization within clusters is also aligned with the CR improvement objective, as Figure~\ref{fig:STD_H} provides a good correlation with CR.    


\subsubsection{Feature selection and clustering metric} 
\begin{figure}
    \centering
    \begin{subfigure}[b]{0.9\linewidth}
        \includegraphics[width=\linewidth]{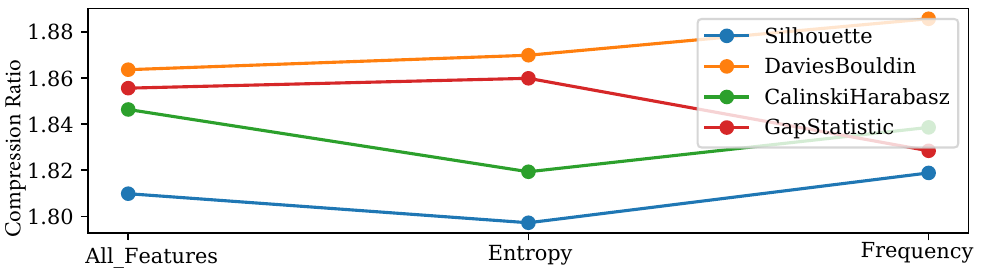}
        \caption{float32 datasets}
        \label{fig:featureSelection32}
    \end{subfigure}
    \begin{subfigure}[b]{0.9\linewidth}
        \includegraphics[width=\linewidth]{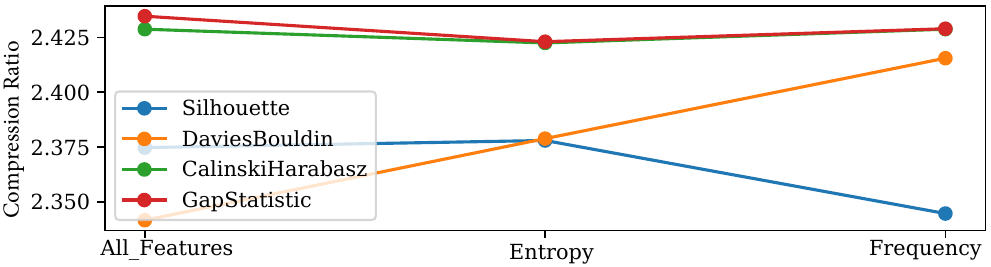}
        \caption{float64 datasets}
        \label{fig:featureSelection64}
    \end{subfigure}
    \caption{The effect of feature and clustering score on CR.}
    \label{fig:featureSelection}
\end{figure}

The first step of \DTT{} is to extract features from each initial byte cluster, later used for clustering. We select a combination of features in Table~\ref{tab:feature}. We tailor the feature selection with a compression pipeline and select features and clustering metrics that provide the highest compression ratio. To do this, datasets are separated into two categories based on their precision: float32 and float64 datasets. 
Then we assessed three feature sets— combined entropy features, frequency, and a combination of both—by measuring their compression ratios. For each choice of feature, four different clustering metrics (Silhouette, Davies-Bouldin, Calinski-Harabasz, and Gap Statistic) are used to cluster the input feature for all datasets. This makes a search space with $12$ different possibilities. We report the GMean compression ratio across all float32 and float64 datasets as shown in Figures~\ref{fig:featureSelection32} and \ref{fig:featureSelection64}, respectively. 

As shown in Figure~\ref{fig:featureSelection}, the optimal feature and clustering metric is different for float32 and float64 datasets. As shown in Figure~\ref{fig:featureSelection32}, the frequency feature, when combined with the Davies-Bouldin metric, provides the highest compression ratio, thus selected for float32 datasets. The selected feature and clustering metric for float64 datasets are combined features and Gap Statistic due to the best compression ratio as shown in Figure~\ref{fig:featureSelection64}. We saw a similar trend in other compression tool and thus analysis for zstd is shown as a representative tool. 


\subsubsection{Clustering algorithm efficiency} 
\begin{figure}
    \centering
    \includegraphics[width=0.9\linewidth]{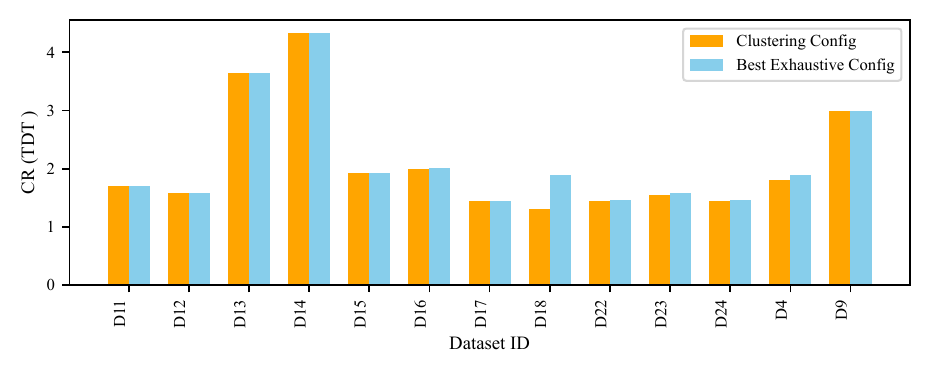}
    \caption{\DTT{} clustering vs exhaustive search using zstd}
    \label{fig:second}
\end{figure}

We discuss the efficiency of \DTT{} clustering from how it picks the optimal point without testing all clustering possibilities and testing all values in the dataset. 
To avoid testing all possible clustering, the second step of \DTT{}  (Algorithm~\ref{alg:clustering}) only tests 4 and 8 possibilities for float32 and float64 datasets. To show the efficiency of the algorithm, we create an exhaustive search algorithm that tests all clustering possibilities for float32 and selects the clustering with the maximum compression ratio as an optimal solution. Then the result of \DTT{} clustering and the exhaustive search algorithm is shown in Figure~\ref{fig:second}. As shown, the \DTT{} clustering step is producing the optimal clustering in 95\% of the datasets and for the remaining it provides a close to optimal clustering. This indicates the efficiency of the clustering step of the \DTT{} algorithm. 

As discussed in Section~\ref{sec:modes}, 30\% of values in a dataset represent characteristics of the full dataset. This approximation reduces the feature extraction time. We conducted two experiments to ensure this approximation does not affect the result. We first redo Figure~\ref{fig:featureSelection} using 30\% datasets, and the outcome is similar in terms of feature and clustering metric selection. Then we additionally used the selected feature and clustering metric from 30\% of datasets and computed the resulting compression ratio and we found that the compression ratio remains the same in 92\% of datasets. Therefore, we select 30\% of values to reduce computation time for clustering.



\subsubsection{Packing Decision} As discussed in Section~\ref{sec:packing}, byte-clusters can be packed to put bytes from the same value next to each other (same-value) or all bytes in the same byte location (same-byte). Since both packing approaches have their advantage, we select the packing approach that provides the highest compression ratio across datasets. We measured both packing approaches across a given clustering and computed the resulting compression ratio across datasets. The compression ratios of datasets were similar in all but two datasets, where same-byte packing performed better, hence the default choice in \DTT.


\subsubsection{Block-size selection} 
As discussed in Section~\ref{sec:blocking}, the block size is selected to use the memory hierarchy of CPU and GPU processors and minimize the effect of load balance and compression ratio. We tune all block sizes between L1 cache and last level cache (LLC), which is L3 in the CPU, to decide about the best block size. For GPU, since the block size is set internally by the tool, we do not specify a block size. To isolate the effect of threading from internal implementation, we use fastlz, which is a single-threaded open-source compression tool. We use blocking to make the tool parallel with the sweep of the block size. 

\begin{figure}
    \centering
    \begin{subfigure}[b]{0.9\linewidth}
        \includegraphics[width=\linewidth]
        {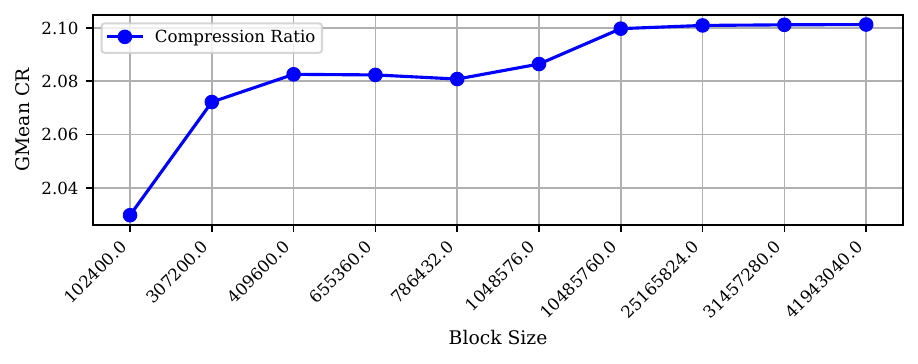}
        \caption{The effect of block size on compression ratio}
        \label{fig:blocksizecompression}
    \end{subfigure}
    \begin{subfigure}[b]{0.9\linewidth}
    \includegraphics[width=\linewidth]
       {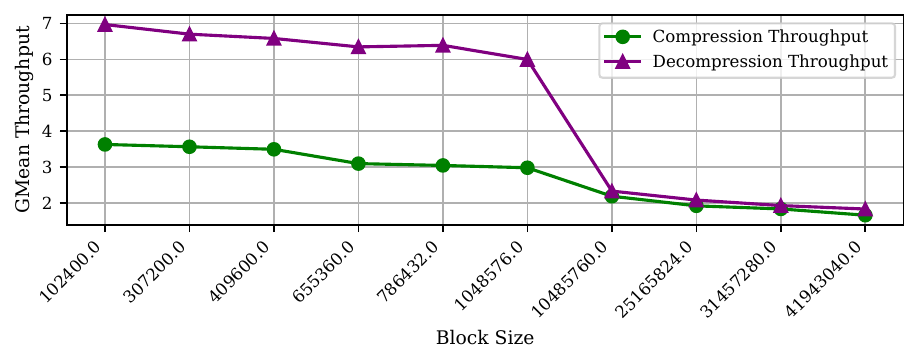}
        \caption{The effect of block size on throughput}
        \label{fig:blocksizethroughput}
    \end{subfigure}
    \caption{Block size trade-off for Fastlz}
    \label{fig:blocksize}
\end{figure}

Figure~\ref{fig:blocksizecompression} shows the effect of block size on compression ratio for the selected block range sweep across all datasets. As shown, the compression ratio for block sizes larger than L1 changes between 2.08--2.1, which is insignificant. Therefore, the goal of blocking is to select a block size that maximizes throughput within the range that has minimal effect on compression ratio. As shown in Figure~\ref{fig:blocksizethroughput}, varying block size drastically affects throughput in both compression and decompression. As shown in Figure~\ref{fig:blocksizethroughput}, between L1 and L2, the throughput is close to the peak throughput. Therefore, we select a block size between L1 and L2 for each compression tool for a good balance between compression ratio and throughput.



\subsubsection{Packing and decomposition interaction with blocking} 
\begin{figure}
    \centering
    \includegraphics[width=1\linewidth , height=4cm]{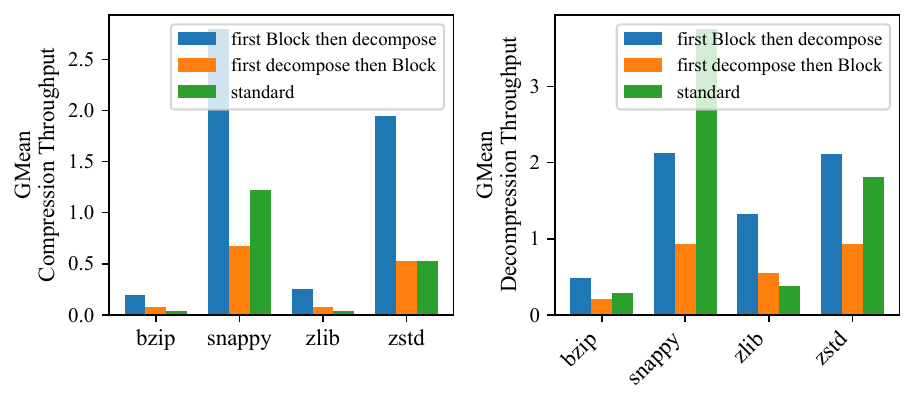}
      
    \caption{Decomposition and blocking order on Throughput (GB/s)}
    \label{fig:blockPackingInterplay}
\end{figure}

As discussed in Section~\ref{sec:blocking}, the order in which blocking and packing are applied does affect throughput. The selected order in \DTT{} is blocking, followed by packing and decomposition. Figure~\ref{fig:blockPackingInterplay} shows the effect of this decision on compression and decompression throughput. As shown, blocking followed by packing leads to a 4.08 and 1.42 improvement over the standard approach in compression and decompression throughput, respectively, while packing followed by blocking results in a 1.22 improvement in compression throughput and a 0.61 slowdown in decompression throughput. The main reason behind this is that blocking can be done without copying overhead, while packing needs explicit copying, as generic compression tools need contiguous memory locations. Therefore, when blocking comes first, the copy overhead packing can be done in parallel.  

\subsection{Discussion and potential extensions}

This section discusses an initial study of the potential extension of \DTT{} for other primitive types such as integers. 
We applied TDT to two integer datasets, Poker-Hand \footnote{\url{https://www.kaggle.com/datasets/camillahorne/poker-hands}} and Covertype\footnote{\url{https://www.kaggle.com/datasets/zsinghrahulk/covertype-forest-cover-types/data}} for two representative compression tools as shown in Table~\ref{tab:CompressionRatioint}. While the study is limited but the result indicates strong potentials in other typed datasets such as integers. \DTT{} provides a CRI between 1.08--1.67$\times$ across these two datasets and compression tools. 

\begin{table}[!ht]
\centering
\resizebox{\linewidth}{!}{%
\begin{tabular}{lcccc}
\toprule
\multirow{2}{*}{\textbf{Dataset}}
  & \multicolumn{2}{c}{\textbf{zstd}}
  & \multicolumn{2}{c}{\textbf{bzip2}} \\
\cmidrule(lr){2-3} \cmidrule(lr){4-5}
  & \textbf{TDT} & \textbf{Standard}
  & \textbf{TDT} & \textbf{Standard} \\
\midrule
\rowcolor{lightgray}
poker-hand (32 bits)
  & 7.25 & 4.35 & 8.29 & 7.26 \\
covertype (64 bits)
  & 32.73 & 23.14 & 40.03 & 36.75 \\
\bottomrule
\end{tabular}%
}
\caption{Compression ratio for integer datasets.}
\label{tab:CompressionRatioint}
\end{table}

\section{Related Work}

\subsection{Compression methods}
We classify lossless compression methods into general and floating-based methods. Common general compression tools treat floating-point data as bytes and compress it using a combination of compression techniques. Some notable compression tools in this category are zstd~\cite{FacebookZstd}, snappy~\cite{GoogleSnappy}, lz4~\cite{YannColletLZ4}, zlib~\cite{gailly2004zlib}, and bzip~\cite{seward1996bzip2}. These compression methods treat floating-point data as bytes and compress them. While effective, these tools do not take advantage of the characteristics of floating-point values.  

Using floating-point standards or domain information to compress floating-point datasets, e.g., in scientific workloads~\cite{lindstrom2017error} and checkpointing data~\cite{son2014data}, is well explored area~\cite{chen2024fcbench}. 
Lossy compression methods such as BUFF~\cite{liu2021decomposed} reduce the precision of the values and apply compression methods. While effective, removing precision is not acceptable in several domains such as medical imaging and machine learning models.
Lossless compression methods often rely on a smooth transition of data in streaming applications and use a prediction-based method to compress the data. The compression methods either use a prediction-based model~\cite{jensen2018modelardb,yu2020two,burtscher2008fpc} or rely on previous values~\cite{pelkonen2015gorilla,liakos2022chimp,li2023elf,afroozeh2023alp} to predict the next value. The smooth pattern in streaming applications creates similar bit patterns in part of floating types, when applying an XOR on consecutive values, resulting in leading or trailing zeros, used for compressing in Gorilla~\cite{pelkonen2015gorilla}, Chimp~\cite{liakos2022chimp}, Elf~\cite{li2023elf}, ALP~\cite{afroozeh2023alp}, and other streaming methods~\cite{ratanaworabhan2006fast}. The other category of floating-point compression uses compression techniques such as prediction-based compression algorithms~\cite{lindstrom2006fast}, dictionary-based methods~\cite{claggett2018spdp}.






\subsection{Data transformation}
Data transformations are commonly applied to various datasets to improve the compression ratios. 
One common data transformation is the Burrows-Wheeler transform (BWT)~\cite{manzini2001analysis} which is typically applied to strings such as in genetic data. BWT is also used in general compression tools such as bzip~\cite{seward1996bzip2}. While efficient for strings, they do not take advantage of floating-point value standards characteristics. 
Another potentially relevant work is the clustering approaches used for compression in database applications such as ~\cite{wandelt2018column}. While the partitioning approach in ~\cite{wandelt2018column} shares common elements, in floating-point datasets, it is also important to ensure high deviation between clusters, necessary for entropy coders. 

Data transformations are applied to floating datasets to improve the compression ratio. 
Brotli~\cite{alakuijala2018brotli} clusters data to group data with a similar nature in one cluster. The clustering scheme in Brotli is coarse and does not differentiate the different entropies of bytes in floating-point datasets.
One of the closely related works is shuffle~\cite{collette2013python} and Bitshuffle~\cite{masui2015compression}. These data transformation approaches reorganized the floating-point data, such as float, at the bit level to create more repetitive patterns in imaging data. While these approached use floating-point properties but they do not consider the effect of bytes together, which differ across datasets. 
\DTT{} transforms floating point datasets based on dataset characteristics and improve compression ratios across applications and compression tools.



\section{Conclusion and Future work}

Lossless compression of floating-point values is critical for several domains. Prior general and floating-point-based compression methods effectively compress floating-point datasets. However, there is still potential to improve compression methods by considering the characteristics of floating-point data. This paper proposes \DTT{}, a technique that uses the non-uniform entropy distribution of different bytes in floating-point datasets to transform the data and create a more uniform dataset. This data transformation is mapped to a clustering problem and implemented in a compression pipeline. As a result, the compression ratio are improved across tools between 1.04--1.17$\times$ and the compression throughput is improved between 1.16--7.57$\times$.
We plan to test this approach for other types of numerical values, such as integers, and evaluate its effectiveness for lossy methods.

\bibliographystyle{ACM-Reference-Format}
\bibliography{ref}

\end{document}